\documentclass[prx,aps,twocolumn,amsfonts,longbibliography,groupedaddress,superscriptaddress,noshowpacs,floatfix]{revtex4}
\usepackage{graphicx}
\usepackage{subfloat}
\usepackage{subfigure}
\usepackage{amssymb}
\usepackage{amsmath}
\usepackage{color}
\usepackage{comment}
\usepackage{multirow}
\usepackage{mathrsfs}
\usepackage{amsbsy}

\newcommand{\Q}{{\pmb{\mathscr{Q}}}}
\newcommand{\n}{{n_{\rm cos}}}
\newcommand{\Qm}{{N_Q}}
\newcommand{\Voph}{{V^{1,p-h}_{\rm cosine}}}
\newcommand{\V}{{V_{\rm cosine}}}

\begin{document}

\title{Glimmers of a Quantum KAM Theorem: Insights from Quantum Quenches in One
  Dimensional Bose Gases}

\author{G. P. Brandino}\author{J.-S. Caux}
\affiliation{Institute for Theoretical Physics, University of Amsterdam,
Science Park 904, 1090 GL Amsterdam, The Netherlands}

\author{R. M. Konik}\email{rmk@bnl.gov}
\affiliation{CMPMS Dept. Bldg 734 Brookhaven National Laboratory, Upton NY 11973 USA}

\begin{abstract}

Real-time dynamics in a quantum many-body system are inherently
complicated and hence difficult to predict.  There are, however, a
special set of systems where these dynamics are theoretically
tractable: integrable models.  Such models
possess non-trivial conserved quantities beyond energy and momentum.
These quantities are believed to control dynamics and thermalization in low dimensional atomic 
gases as well as in quantum spin chains.
But what happens when the special symmetries leading to the existence of the extra conserved quantities are broken?  Is
there any memory of the quantities if the breaking is weak? Here, in the presence of weak integrability breaking,
we show that it is possible to construct residual quasi-conserved quantities, so providing
a quantum analog to the KAM theorem and its attendant Nekhoreshev estimates.
We demonstrate this construction explicitly in the context of quantum quenches
in one-dimensional Bose gases and argue that these quasi-conserved
quantities can be probed experimentally.

\end{abstract}
\maketitle

A milestone in the dynamics of classical many-body systems is the
Kolmogorov-Arnold-Moser (KAM) theory \cite{kam}.  Generically, classical many-body systems exhibit
chaotic behaviour -- that is to say, giving the bodies of such systems slightly different
initial positions and velocities results in the bodies following radically different trajectories.
An exception to this rule is made for a special set of systems termed
{\it integrable} which possess conserved
quantities beyond energy and momentum.  The existence of these conserved quantities promises
the availability of a set of action-angle $\{p_i,q_i\}$ whose
action variables are constants of motion.  In such variables the Hamiltonian, $H$, is solely a function
of $\{p_i\}$, and Hamilton's equations of motion become particularly simple:
\begin{equation}
\dot q_i = \frac{\partial H}{\partial p_i}, ~~~ \dot p_i = 0.
\end{equation}
Trajectories of bodies in integrable systems
are not sensitive to initial conditions, but instead lie on invariant tori in phase space described by frequencies $\{\omega_i\}$ parameterizing
solutions to the equations of motion: $\dot q_i = \omega_i$.
However integrable systems and their attendant simple behaviour are comparatively rare.  And so the question
arises what can one expect with a system which is merely close to being integrable.
Is the motion of bodies in this system chaotic?  Or is there some influence on the system's dynamic from being close
to an integrable point?  One answer to this question is given by the KAM theorem.  It tells us that if we weakly
perturb a classical integrable system, we do not immediately transit to completely chaotic dynamics, but rather
see a smooth crossover.  
Specifically, the KAM theorem promises that a subset of the solutions $\{\omega_i\}$ survive under a sufficiently
small perturbation, $\epsilon H_{pert}(p_i,q_i)$, provided their frequencies are sufficiently irrational.

What of quantum analogs to the KAM theorem?  There is tremendous interest 
\cite{fior_muss,quench_action,rigol,rigol1,caza,caza1,NRG4,caux,caux1,n_andrei,zotos,sirker,rossini,prosen,fag_cal_ess,fag_cal_ess1,fag_cal_ess2,tg_quench1,tg_quench2,fail_gge1,fail_gge2,fail_gge3,xxz_quench,poszgay_fail,muss_tak,gge_ft,caux_complete_gge,prosen_semilocal,prosen_semilocal,caux_complete_gge}
in the role exotic conserved quantities play in the dynamics of low dimensional systems.  This interest 
\cite{rigol,rigol1,caza,caza1,NRG4,caux,caux1,n_andrei,tg_quench1,tg_quench2} arises in the context of one dimensional (1D) Bose gases 
from the ability to manipulate isolated gases and observe their relaxation in closed surroundings, 
both when the gases are near integrable points \cite{weiss,schmiedmayer,schmiedmayer1} as well as far away \cite{nagerl}.
In the context of quantum spin chains \cite{zotos,sirker,rossini,prosen,fag_cal_ess,fag_cal_ess1,fag_cal_ess2}, it comes about 
from the wish to understand related thermalization questions as well as whether integrable systems can sustain ballistic transport.
It also appears in the burgeoning field of many-body localization \cite{altshuler,vadim} of disordered interacting systems and associated attempts to construct
sequences of conserved charges in what one would traditionally consider a non-integrable setting \cite{serbyn,imbrie}.  

To understand crossover behavior
arising from integrability breaking, both indirect measures
such as level spacing statistics \cite{berry-tabor,NRG2,rigol_ls1,rigol_ls2} as well as studies of systems in their quasi-classical 
limit using such tools as the semi-classical eigenfunction hypothesis \cite{percival,berry,voros} are oft employed.
Such behavior is often phrased in terms of pre-thermalization
\cite{berges,kehrein,kehrein1,kollar,ess_rob,pre_njr} or pre-relaxation
plateaus \cite{fagotti1,fagotti2,fagotti3},
where a system's observables, in relaxing from some non-equilibrium
initial state, remain nearly constant over 
some finite time interval before decaying to their final equilibrium value.
Such plateaus have been argued to be controlled by the remnants of the conserved quantities of the nearby integrable system \cite{kollar,ess_rob}.

In this work we go beyond this and show that in finite systems it is possible to construct an infinite sequence of nearly conserved local operators, 
$\{\Q_i\}^\infty_{i=1}$, in the presence of a perturbation that weakly breaks integrability,
\begin{equation}
H = H_{\rm integrable} + \epsilon \Phi_{\rm perturbation}.
\end{equation}
We will show that this near-conservation is {\it good for all times}.
The $\Q_i$ are conserved in the sense that if we consider
a (non-eigen)state, $|s\rangle$, with average energy per particle
$E=\langle s|H|s\rangle/N$ less than some bound $\Lambda (N_Q)$, then 
\begin{equation}
\partial_t \langle s| \Q_i(t)|s\rangle < \delta(\epsilon ,N_Q ),
\end{equation}
for all times where $\delta (\epsilon ,N_Q )$ can be made arbitrarily small.
These conserved operators are constructed as finite linear
combinations (length $N_Q$) involving the charges $\{\hat Q_i\}^\infty_i$ of
the unperturbed Hamiltonian, $H_{\rm integrable}$:
\begin{equation}
{\Q}_i(N_Q) = \sum_{j=1}^{N_Q} a_{i,j} \hat Q_{j+iN_Q} .
\end{equation}
The quality of this conservation can be controlled (i.e. $\Lambda$ can be made larger and $\delta$ smaller)
by adjusting how many, $N_Q$, of the charges, $\hat Q_i$, appear in the linear combinations.  

\begin{figure*}[t]\label{Fig1}
\center{\includegraphics[width=.8\textwidth]{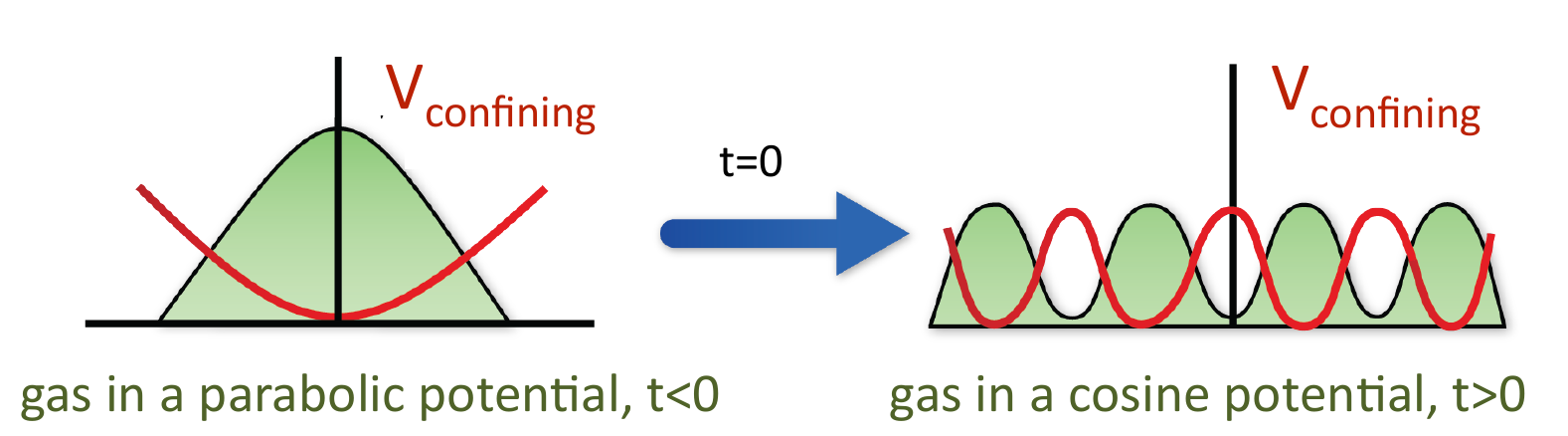}}
\caption{{\bf Quench protocol}: We prepare the one dimensional Bose gas
in its ground state in a harmonic trap.  At time t=0 we release the gas into
a cosine potential and track the subsequent dynamics.  The shaded green regions are illustrations of the
equilibrium density profiles of the gas in the presence of the confining potentials.}
\end{figure*}

Our construction is akin not so much to the KAM theorem, but to what are known as Nekhoroshev estimates \cite{nekhoroshev}
inasmuch as the charges $\Q$ we construct are nearly conserved on the entirety of the low energy Hilbert space.
While the KAM theorem promises that some subset of solutions
of the equations of motion survive a perturbation {\it and} remain ``close'' to their integrable counterparts for all
time, the Nekhoroshev estimates tell us that {\it all} solutions remain close to their integrable counterparts 
in the sense that 
\begin{equation}\label{Nek1}
|p_i(t)-p_i(0)| < P_{*}\epsilon^{\frac{1}{2N}},
\end{equation}
for exponentially long times:
\begin{equation}\label{Nek2}
t < T_{*} e^{\big(\frac{a}{\epsilon}\big)^\frac{1}{2N}},
\end{equation}
where here $P_{*}$, $T_{*}$, and $a$ are constants and $N$ is the number of degrees of freedom the system has \cite{nekhoroshev}.

\begin{figure}[b]
\begin{center}
\includegraphics[width=0.4\textwidth]{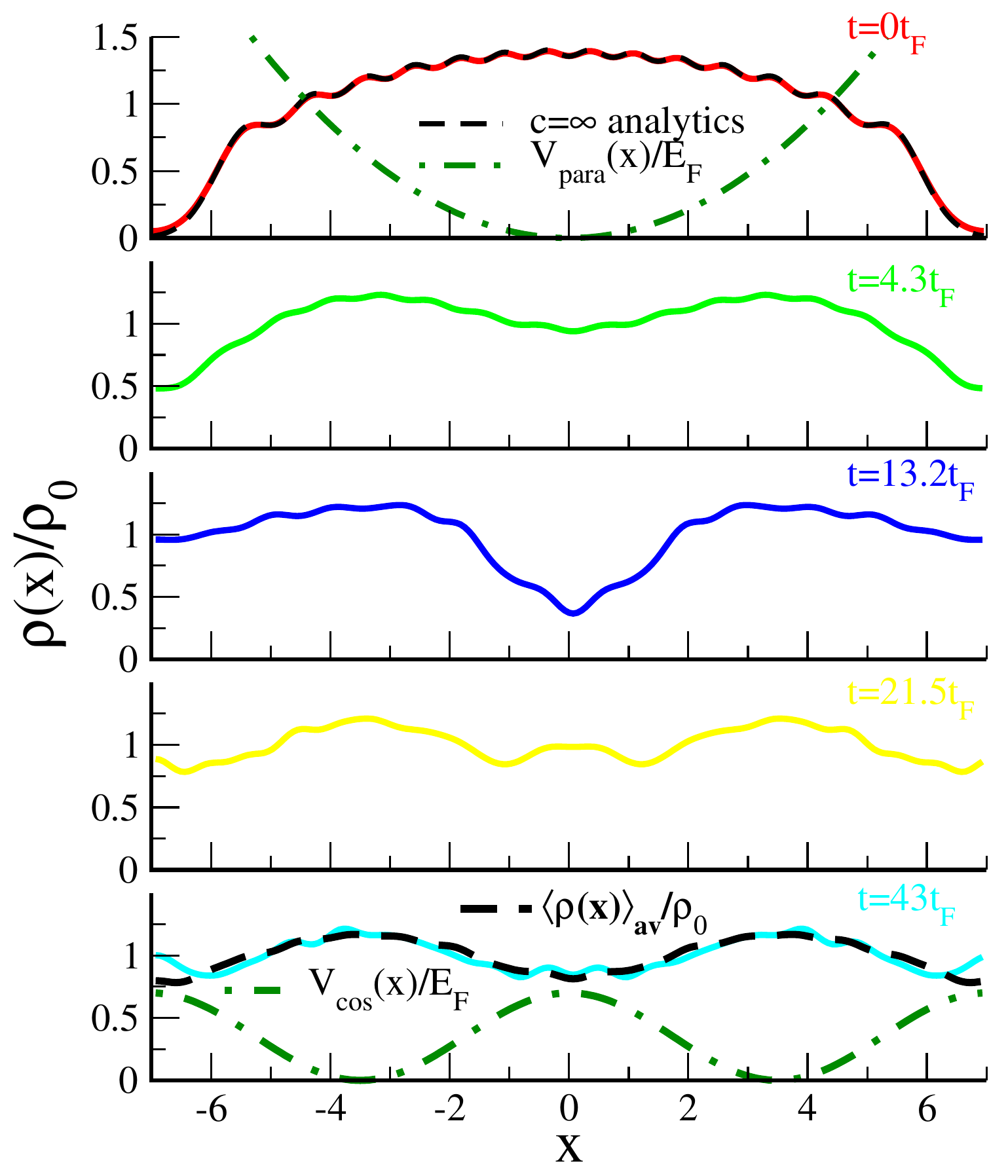} 
\end{center}
\caption{\label{fig:timeev_rho}The density profile of the
gas at selected times post-quench as computed with the NRG. Here this time dependence is computed
after releasing a $N=L=14$, $c=7200$ gas prepared in a parabolic
potential with $m\omega_0^2L^2/2E_F=10.36$ (shown with a green dashed line in the $t=0t_F$ frame, $t_F=1/E_F$, $E_F=k_F^2/(2m)$, and $k_F=\pi(N-1)/L$) into a cosine potential
$V_{\rm cosine}(x)/E_F=0.35(\cos(\frac{4\pi}{L}x)+1)$ (plotted with a dashed line in the $t=43t_F$ frame). In the $t=0$ frame, we show the density
profile as computed analytically in the hardcore limit (see Appendix A2).  Using the NRG we can run the time evolution as far out as $t=85t_F$
before dephasing exceeds $1\%$.  We see however by $t=43t_F$ the gas' density profile has already come close to its long time average (black
dashed line in the final panel).}
\end{figure}

While general, we perform this construction in the context of quantum quenches in one dimensional (1D) Bose gases.  
This setting is particularly appropriate as it is the experimental study of quantum quenches in these gases \cite{weiss} that has
led to tremendous interest in the role of exotic conserved quantities in quantum dynamics.  Quenches are moreover directly relevant to understanding these
experiments.  Because of the one-body potentials that trap the gases, they can be at most approximately integrable.  Thus the construction
of a quantum version of the KAM theorem and its variants can only help yield insights into the dynamics of these gases in their experimental
settings.

\section{Quantum quench dynamics in 1D Bose gases}

To set the scene, we first describe the quantum quench in a 1D Bose gas as 
described by the Lieb-Liniger model \cite{ll}.  The Lieb-Liniger model is believed to provide an excellent description of a 1D Bose gas \cite{review}.
In the absence of external (trapping) one-body terms, it is integrable with an infinite number of conserved operators, $\{\hat Q_i\}$.
It's Hamiltonian with the addition of a one body potential, $V(x)$, is given by 
\begin{equation}
H = -\frac{\hbar^2}{2m}\sum^N_{j=1}\frac{\partial^2}{\partial x_j^2}+2c\sum_{\langle i,j\rangle}\delta(x_i-x_j) + \sum_i V(x_i).
\end{equation}
The type of quantum quench we will study is found in preparing the gas
on a ring of length $L$ in the ground state of a parabolic trap \cite{NRG4,tg_quench1,tg_quench2}, i.e.
$V(x)=\frac{1}{2}m\omega^2x^2$, then at time $t=0$, releasing the gas from the parabolic trap into
a one-body cosine potential, $V(x)=A\cos(2\pi \n x/L)$, and observing the subsequent dynamics of the gas.
This quench protocol is illustrated in Fig. 1.

This form of the Hamiltonian, an integrable model together with an integrability breaking perturbation, allows
us to determine the ground and excited states of the model pre- and post-quench through a numerical renormalization group (NRG) 
designed precisely to attack such problems \cite{NRG,NRG2,NRG3,NRG4} together with a set of routines known as ABACUS that 
allow {\it numerically exact} computation of matrix elements of operators in the Lieb-Liniger model \cite{caux_abacus}.  In turn,
this gives us access to the post-quench dynamics of the gas.
In particular we employ an NRG able
to study perturbations of integrable and conformal continuum field theories.  This 
approach, as it is an extension of a methodology known as the truncated conformal spectrum
approach \cite{YZ,YZ1}, has been primarily used to study perturbations of relativistic field theories \cite{NRG,NRG2,NRG3}, but
has recently been applied to the Lieb-Liniger model perturbed by a one-body potential \cite{NRG4}, the problem
at hand.  The NRG uses the eigenstates of the Lieb-Liniger model as a computational basis.  Because this
basis accounts for the interactions of the Bose gas particles with one another, this numerical method builds
in the strong correlations present in the problem right at the start.  We discuss details of this method
in Appendix A1.

In Fig. 2 we show the time evolution of the gas after the quench.  At time $t=0$ we see the density profile of the gas 
in the ground state of the parabolic potential.  After quenching the potential to a cosine, the gas moves away from the center,
oscillates a number of times before settling into the minima of the cosine.  This occurs at times of the order of $t=50t_F$ -
we are able to run the simulation out to times of $t=80t_F$ (here
$t_F=1/E_F$ where $E_F=k_F^2/(2m)$ and $k_F=\pi(N-1)/L$).

While we are able to compute the dynamics of such observables as the density and the momentum distribution function, the
key to the work in this paper will be our ability to compute the dynamics of the (formerly) conserved Lieb-Liniger charges, $\hat Q_i$.
Our numerical approach makes this extremely simple because of our use of the eigenstates of the integrable Lieb-Liniger model
as a basis.  
Each Lieb-Liniger state of an $N$-particle gas $|\psi\rangle_{LL}$ is characterized by $N$-rapidities, $\lambda_i,~i=1,\ldots,N$, which
should be thought of as, more or less, the momenta of the gas's particles.  
These rapidities determine the action of the conserved operators on the Lieb-Liniger states.  For
example both the energy, $E=\hat Q_2$ and momentum, $P=\hat Q_1$, operators act on $|\psi\rangle_{LL}$
via,
\begin{equation}
E|\psi\rangle_{LL} = \sum^N_{i=1} \lambda_{i}^2|\psi\rangle_{LL}; ~~~ P|\psi\rangle_{LL} = \sum^N_{i=1} \lambda_{i}|\psi\rangle_{LL}.
\end{equation}
The action of all of the higher non-trivial charges, $\hat Q_n$, $n=3,4,5,\cdots$ in the Lieb-Liniger model
are simply higher power sums of the same rapidities:
\begin{equation}
\hat Q_n|\psi\rangle_{LL} = \sum^N_{i=1} \lambda_{i}^n|\psi\rangle_{LL}.
\end{equation}
While the actual expression of the charges in terms of the Bose field operators is complicated and unwieldy \cite{korepin}, 
the action of the charges on the Lieb-Liniger eigenstates turns out to be extremely simple.  This will be crucial in facilitating
our construction of effective $\Q$'s.

\section{Construction of conserved quantities in the Bose gas post-quench}

\begin{figure*}[t]
\begin{center}
\includegraphics[width=1\textwidth]{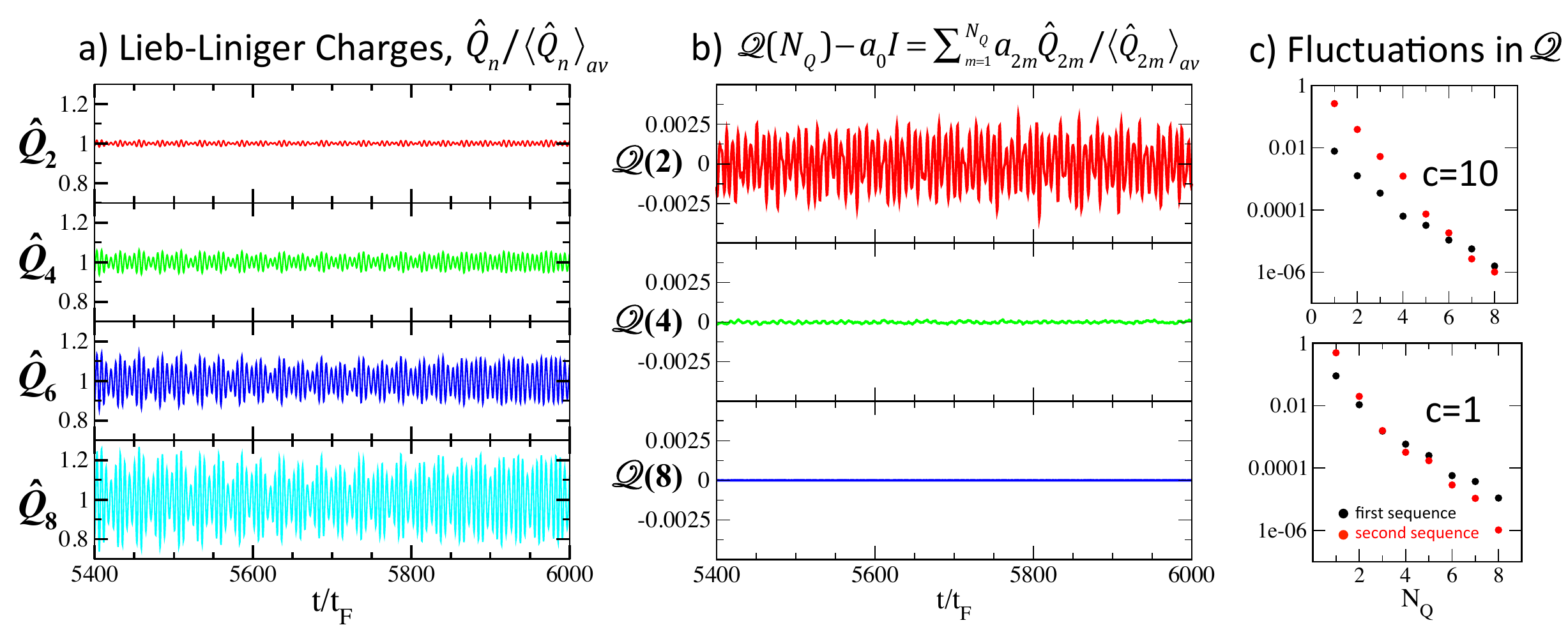} 
\end{center}
\caption{\label{fig:timeev}a) The post-quench time evolution of the Lieb-Liniger charges normalized by their mean value
as described in the text.  Here the time
dependence is computed after releasing a $N=L=8$, $c=10$ gas prepared in a parabolic potential of strength $m\omega_0^2L^2/2E_F=3.24$ 
into a cosine potential $\cos(\frac{4\pi}{L}x)$.  We show this behavior at late times (for details of how long we
can run the simulation, see Appendix A1).
b) The post-quench time evolution of a sequence of 
effective charges, $\Q(N_Q) = \sum_{m=1}^{N_Q}a_{2m}\hat Q_{2m}$, for $N_Q=2,4,$ and $8$. 
c) Top: The standard deviation of the fluctuations of two sequences of effective charges $\Q$.  We build the first sequence (in black)
using linear combinations of the charges $\{\hat Q_{2m}\}_{m=1}^{m=8}$, while the second sequence (in red) is formed with
the next eight Lieb-Liniger charges, i.e. $\{\hat Q_{2m}\}_{m=9}^{m=16}$.  Bottom: We show the fluctuations of the two effective charges
built following the quench of a $c=1$ gas prepared in a parabolic trap of strength, $m\omega_0^2L^2/2E_F=0.13$, and released into the same cosine
potential, $\cos(\frac{4\pi}{L}x)$.}
\end{figure*}

We now turn to the core of the paper.  We have shown in the previous section that we can
describe the temporal dynamics of various quantities post-quench.  In that section we specifically considered the density profile
of the gas after release into the cosine potential.  We now consider the time evolution of the Lieb-Liniger charges.  They
are of course not conserved and so their evolution will be non-trivial.  We however show that one can construct linear combinations
of the Lieb-Liniger charges whose expectation values are nearly time invariant under unitary evolution by the post-quench Hamiltonian.  The quality
of this time invariance can be controlled by allowing more charges in the linear combination.  Moreover
we show that these linear combinations of charge are not merely time invariant with respect to the particular initial condition created
in the quench protocol, but as operators acting on the low energy Hilbert space.

\begin{figure}[b]
\begin{center}
\includegraphics[width=0.4\textwidth]{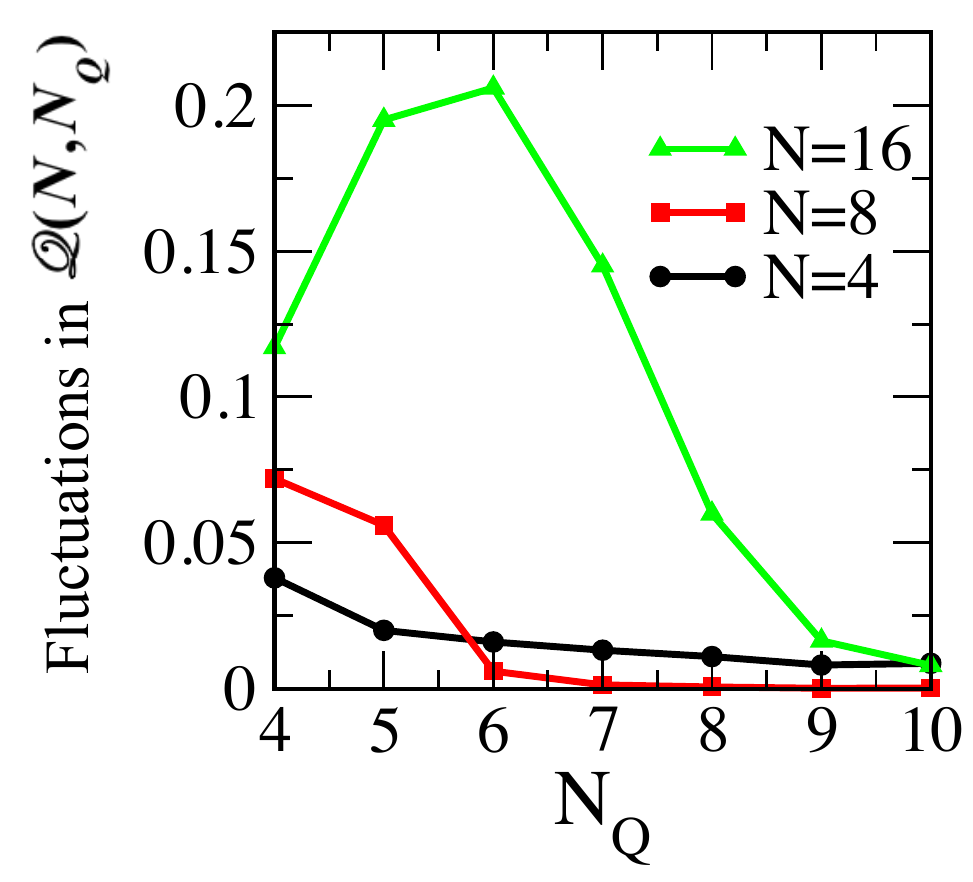}
\end{center}
\caption{\label{flucts_vs_NQ} We plot the fluctuations in time for $\Q$ 
as a function of $N_Q$ for $N=L=4,8,$ and $16$ for a quench
from a parabolic potential of strength $m\omega_0^2L^2/2E_F =2.33$ to a cosine of amplitude $V_{cos}(x)=0.26E_F\cos(2\pi x/L)$.
We do so using the charges as constructed at $c=\infty$  as discussed in Appendix B as a partial demonstration that
such charges work well at finite c.}
\end{figure}

We begin by first considering the time evolution of the individual Lieb-Liniger charges themselves.  We plot this evolution for the first
four Lieb-Liniger charges in Fig. \ref{fig:timeev} for a gas with $N=L=8$ and $c=10$.   In plotting the time evolution we have normalized each charge 
to its mean value post-quench so that all of the charges fluctuate about 1.  
The mean value of the unnormalized n-th charge, given by,
\begin{equation}
\langle \hat Q_n \rangle_{\rm av} = \frac{1}{T}\int^T_0 \langle \hat Q_n(t)\rangle,
\end{equation}
where $T$ is the time out to which we can track the evolution,
grows rapidly with n
as the charge's action on a Lieb-Liniger eigenstate $|s\rangle =|\lambda_1,\cdots,\lambda_N\rangle$ is a power sum of the rapidities
$\{\lambda_i\}_{i=1}^N$, i.e. $\langle s|\hat Q_n|s\rangle = \sum_i \lambda_i^n$. 
We see from Fig. \ref{fig:timeev} that even after normalization, the size of the oscillations increases with $n$.

\begin{figure*}[t]
\begin{center}
\includegraphics[width=1\textwidth]{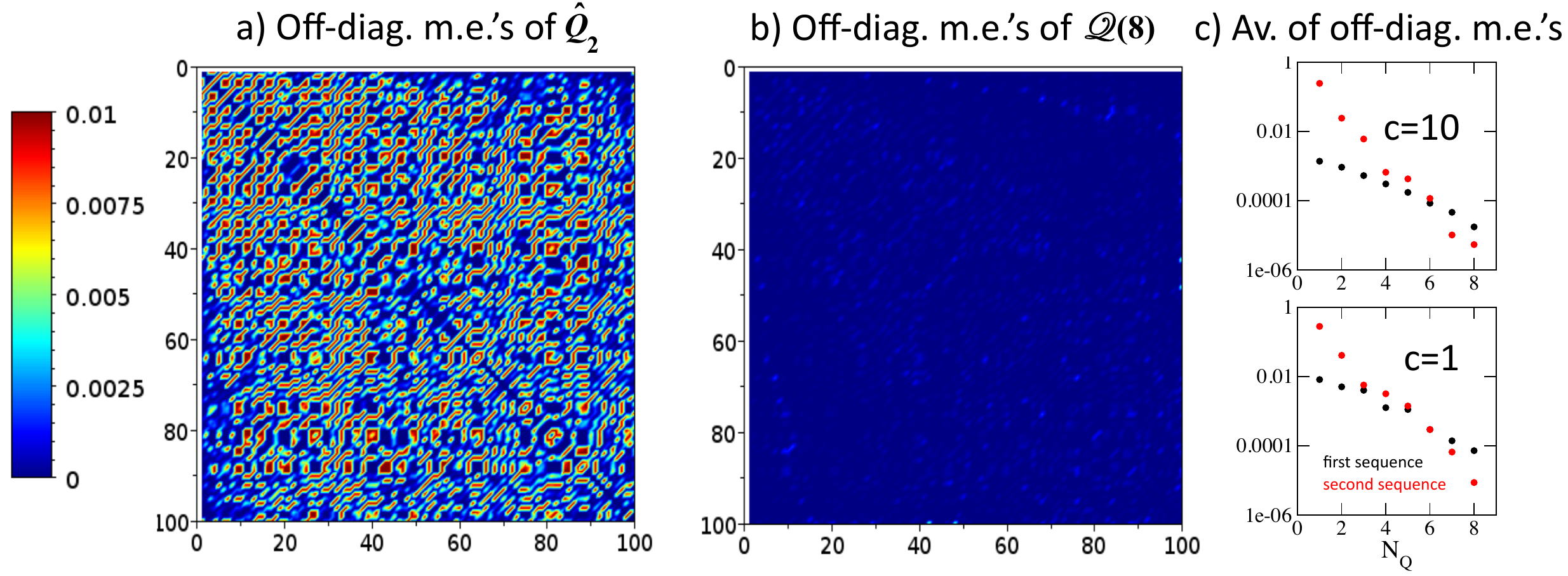}
\end{center}
\caption{\label{off-diag-mes}a) We plot the intensity of the off-diagonal matrix elements of $\hat Q_2$, comparing it to b) 
the off diagonal m.e.'s of $\Q (8)$ for the quench of the $c=1$ gas discussed in Fig. 3c. 
c) We plot the average size of the off-diagonal matrix elements of two sequences of effective charges $\Q(N_Q)$, in black is
the sequence constructed from $\hat Q_{2m}$, $m=1,\cdots, 8$, while in red is the sequence constructed from
$\hat Q_{2m}$, $m=9,\cdots, 16$.  We show this for both the $c=1$ (same quench as in a) and b)) and the $c=10$ case (same quench as
described in Fig. 3a)-c).  }
\end{figure*}

We now consider linear combinations of the Lieb-Liniger charges of the form:
\begin{equation}
\Q (N_Q)  = a_0{\rm I} + \sum_{i=1}^{N_Q} \frac{a_i}{\langle \hat Q_{2i}\rangle_{\rm av}} \hat Q_{2i}; ~~~1=\sum_{i=1}^{N_Q} |a_i|^2 ,
\end{equation}
where we choose the constant $a_0$ such that the mean value of $\Q (N_Q)$ is about 0 and the remaining constants $a_i$\cite{coeff_fn} such
that the fluctuations in $\Q (N_Q)$ are minimized.

We plot the time evolution for a $c=10$ gas of these effective charges in panel b) of Fig. \ref{fig:timeev} for three
different values of $N_Q$, the number of charges in the linear combination.
In panel c) we plot the fluctuations of this charge as a function of $N_Q$. 
We see that these fluctuations drop exponential with $N_Q$.  (On the basis of an error analysis in our
numerics, we would put a numerically induced floor of $10^{-6}$ to $10^{-7}$ on the fluctuations in $\Q$ -- 
see end of Appendix A1a.)
In the bottom part of panel c) we do the same for a quench involving a $c=1$ gas.
In order to be sure that we
are not simply reconstructing the post-quench Hamiltonian as some linear combination of the Lieb-Liniger charges, 
in both cases ($c=10,1$), we demonstrate we can construct
simultaneous multiple effective charges.  In panel c) we show that the fluctuations of a second effective charge
built as a linear combination of charges drawn from $\{\hat Q_{2n}\}_{n=9}^{16}$ also die off exponentially.

This exponential dependence in $N_Q$ is possible to understand at large $c$.  To do so, we write the initial condition of the gas in 
terms of post-quench cosine eigenstates: 
$|\psi_{GS}\rangle = \sum_\alpha c_\alpha |\psi_{\alpha,{\rm cos}}\rangle$.
With the initial condition as above, the time dependence of the charge takes the form:
\begin{eqnarray}
\Q (t) &=& \sum_{\alpha\beta} c^*_\alpha c_\beta \langle \psi_{\alpha,{\rm cos}}|\Q (t)|\psi_{\beta,{\rm cos}}\rangle\cr\cr
&=& \sum_{\alpha\beta} c^*_\alpha c_\beta e^{-i(E_\beta-E_\alpha)t}\langle \psi_{\alpha,{\rm cos}}|\Q |\psi_{\beta,{\rm cos}}\rangle.
\end{eqnarray}
We demonstrate in Appendix B1 that each Lieb-Liniger charge forming $\Q$ 
zeroes a shell of matrix elements $\langle \psi_{\alpha,cos}|\Q |\psi_{\beta,cos}\rangle$, $\alpha \neq \beta$, in the above sum. 
As $N_Q$ increases, more and more of these
matrix elements are zeroed out.  For relatively weak cosine potentials, the total weight, $W_{elim}$ of the $|c_\beta c_{\alpha}|^2$'s
whose matrix elements are zeroed out is
\begin{equation}\label{weakA}
W_{elim}  \approx 1 - 2\frac{e^{-\Lambda(N_Q)^2}}{\sqrt{\pi}}\sum^{N-1}_{n=0}\frac{2^n\Lambda(N_Q)^{2n-1}}{n!}
\end{equation}
with $\Lambda(N_Q) = (2\pi (N_Q-2)/(L\sqrt{m\omega_0})$.
We then see the weight that is not zeroed out and so can contribute to $\Q$'s temporal fluctuations goes as $e^{-(\Lambda(N_Q)/m\omega_0)^2}$.  
We see from this that it becomes harder to construct quasi-stationary, $\Q$'s, as the system size, $L$, is increased.   This
is confirmed in Fig. \ref{flucts_vs_NQ}, where we compare $\Q$'s constructed at different $N=L$.  We see that the point
where the fluctuations become exponentially small goes as $N_Q=L$.

For large amplitude $A$ cosine potentials, the temporal fluctuations die off much more slowly with $N_Q$:
\begin{equation}\label{strongA}
W_{elim} \sim \bigg(\frac{N_Q}{N_A}\bigg)^N, ~~ N_A = \frac{\sqrt{2mA}L}{2\pi}.
\end{equation}
In this latter case, essentially the number of non-zero matrix elements of $\Q(t)$ proliferate, making a construction
where it is nearly time invariant much more difficult.

So far we have only demonstrated that we can construct charges $\Q$ as linear 
combinations of the original Lieb-Liniger charges, $\hat Q_n$,
whose time fluctuations can be made arbitrarily small supposing we start the system in a 
specific initial condition, $|\psi_{GS,para}\rangle$.
However we now demonstrate that these charges are quasi-conserved not just relative to a 
specific initial state, but as operators,
at least when projected onto the low energy post-quench Hilbert space.

To do so we compute the off-diagonal matrix elements in Fig. \ref{off-diag-mes} of one of the two $\Q$'s we have
constructed (the one constructed with Lieb-Liniger charges, $\hat Q_2,\cdots,\hat Q_{16}$) 
relative to the basis of the low-lying energy eigenstates of the post-quench Hamiltonian.
These matrix elements are plotted in Fig. \ref{off-diag-mes}.  In the rightmost panel
we display the off-diagonal matrix elements of $\hat Q_2$ (normalized as described previously) to set the scale of how large these
matrix elements are for the individual Lieb-Liniger charges.  In the middle panel we then plot the matrix elements of $\Q (8)$.
We see that most of the previous non-zero matrix elements of $\hat Q_2$ are now dramatically
reduced.  We quantify this disappearance in panel c) of Fig. \ref{off-diag-mes}.  
There we present
the average magnitude of the off-diagonal matrix elements as a function of $N_Q$.  We present data for both effective charges
considered in Fig. \ref{fig:timeev} for both values of $c=1,10$.  We see in all cases the size of these matrix elements drops exponentially
in $N_Q$.  Roughly speaking, if the average energy per particle of two
distinct states, $|s\rangle, |s'\rangle$, is less than $\Lambda (N_Q)$, then $\langle s|\Q | s'\rangle$ will be exponentially small.
We conclude that the $\Q$'s are then nearly conserved as operators.  This conclusion is supported by an analytic
construction of the $\Q$'s that we present in Appendix B.

\section{Discussion}

In this paper we have found a construction of quasi-conserved operators
as linear combinations of the Lieb-Liniger conserved charges.  In this
construction, the linear combinations are chosen to minimize the temporal fluctuations of the charge
upon quenching the gas from a one-body parabolic potential to a cosine potential.  Despite this
minimization being done for a particular quench protocol, the conservation of the charge occurs
at the operator level.  Specifically, off-diagonal matrix elements of the charges are small.
We demonstrated that both post-quench temporal fluctuations and the off-diagonal matrix elements
can be made exponentially small in the number of charges, $N_Q$, in the linear combination.
We have supported this construction by demonstrating an equivalent analytic construction of these
charges (Appendix B).

In this analytic construction of effective charges we demonstrate why certain linear combinations
of the original Lieb-Liniger charges act as effective conserved quantities at low energies.  This construction
works by finding linear combinations that zero out off-diagonal matrix elements at a given order in the
effective charge $\Q(t)$ written as a power series in time $t$.  We show in particular that a matrix element zeroed out
at a given order in $t$ remains zero to a much higher order in general, thus providing an explanation why our construction
appears so robust.  We stress that this construction uses in no fashion the fact that there does exist a set of exact
conserved charges at $c=\infty$ (namely the occupation numbers belonging to the single particle
states of a cosine potential).  However to reassure the reader that our $c=\infty$ construction is not accidentally in fact
constructing these occupation numbers, we demonstrate that the charges we analytically construct at $c=\infty$ work
at finite $c$ as well.  In Fig. 6 we plot the temporal fluctuations of the effective charges as 
a function of $N_Q$ so analytically constructed
but for the $c=1$ and $c=10$ quenches described in Fig. 3.  While we see the temporal fluctuations of these analytical $c=\infty$ charges
are larger than those numerically constructed at a given $c$ (compare Fig. 3c), we nonetheless see that the fluctuations in the $c=\infty$
charges die off exponentially with $N_Q$.  A similar conclusion can be seen in our study of the temporal fluctuations
of $\Q$ as a function of $N$ and $N_Q$ in Fig. 4 where we again have used the $c=\infty$ $\Q$ -- although here, for
the $N=4$ data, one can see that the fluctuations for the analytic $\Q$ have a comparatively large floor.
All together this gives us confidence that our $c=\infty$ construction is accurately capturing
the essence of the numerical construction of $\Q$ at finite c.

\begin{figure}[t]
\begin{center}
\includegraphics[width=0.4\textwidth]{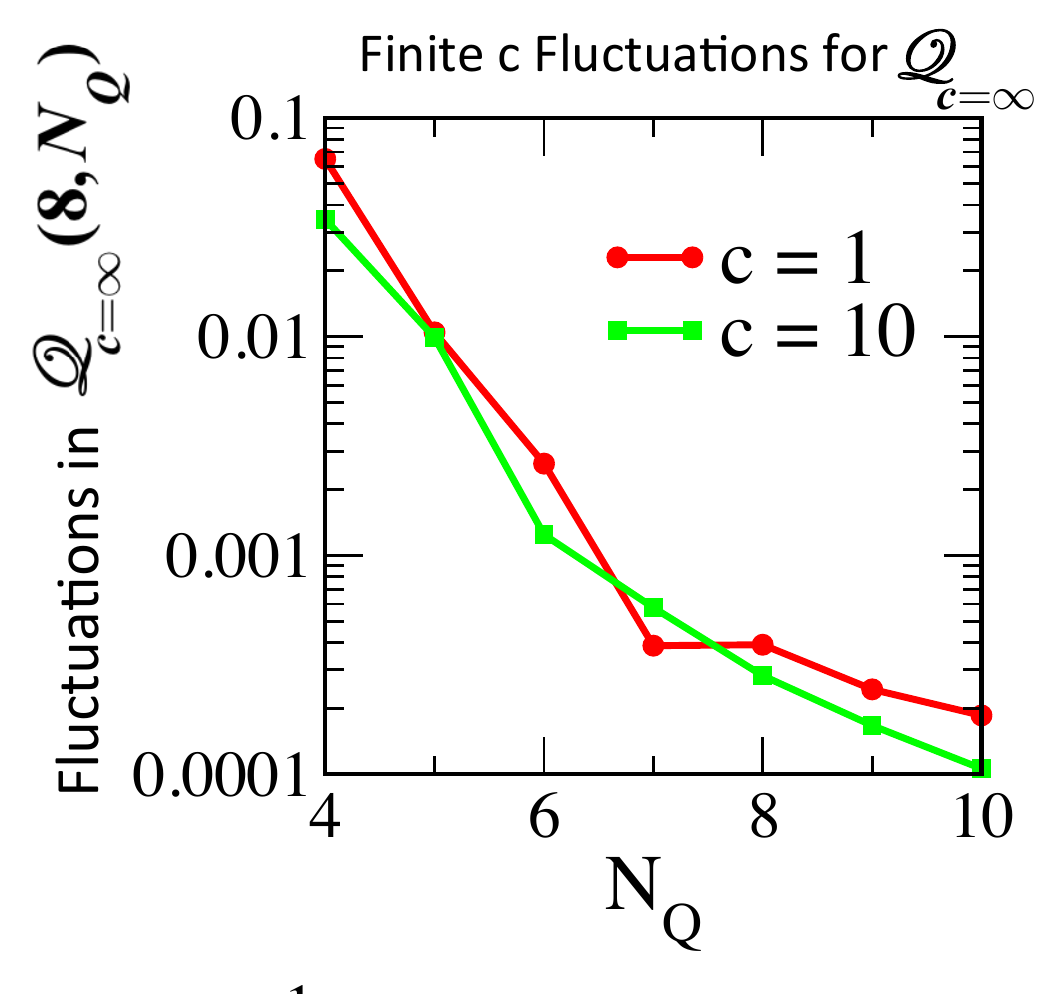}
\end{center}
\caption{\label{fluct_hc} We demonstrate that the effective charges constructed analytically at $c=\infty$
as described in detail in Appendix B have suppressed temporal fluctuations for quenches with finite $c=1,10$.}  
\end{figure}

We are able to in fact extend the analytic computation described in Appendix B to the finite $c$ case.  The primary difference between
the construction of $\Q$ at $c=\infty$ and $c$ finite is the need to take into account that the density operator can connect
states differing by more than one particle-hole pair.  However these higher particle-hole process are suppressed in powers of $1/c$,
with $c$ the interaction strength.  This means that we have a control parameter in our finite $c$ analytic computation of $\Q$
where if we ignore processes involving $n-$particle-hole pairs, the error we make is only $c^{-n}$.  This in part explains why our
$c=\infty$ construction of $\Q$ still is conserved at $c=10$.  It is however somewhat surprising that our $c=\infty$
construction of $\Q$ works as well as it does (as evidenced in Fig. \ref{fluct_hc}) for $c=1$.  This suggests that higher particle-hole
processes, at least for quenches whose dynamics are restricted to the low energy post-quench Hilbert space, are unimportant.

In the introduction to this paper, we have billed these constructions as being quantum equivalents
to the quantum KAM theorem and its counterparts such as the Nekhoreshev estimates.  
There are some similarities in the consequences of our constructions as well
as some dissimilarities.  Nekhoreshev estimates tell us
that the values of the classical action variables in the face of a small non-integrable perturbation change only very slowly in time,
as controlled by both the size of the perturbation and the number of degrees of freedom (see Eqns. \ref{Nek1}
and \ref{Nek2}).

For the quantum case, we see something analogous but with certain differences.  These differences arise both
because we are forming linear combinations of the originally conserved charges, and because of how in our
construction we segregate portions of the quantum phase (Hilbert) space.  Nekhoreshev estimates apply to the entire
phase space of the weakly perturbed model (i.e. Eqn. \ref{Nek1} is good for any $p_i(t=0)$).  In contrast, in our constructions, 
the approximate time invariance of the charge is restricted to a portion of the low energy Hilbert space as marked
by the integer $N_{max}$ (this low energy Hilbert space is defined by states where none of the particles in the state
have momenta greater than $k_{max}=2\pi N_{max}/L$).
While we can make $N_{max}$ as large as we want (provided we are willing to make $N_Q$ correspondingly
large), we cannot take it to be infinite.

Another difference between the two constructions is the role played by the strength of the integrability breaking
perturbation.  Here the Nekhoreshev estimates provide a bound on the temporal variation of
the original action variable going as a fractional power (a function of the system's degrees of freedom)
of the strength of the perturbation.  We, in contrast, can construct effective charges, $\Q$, whose temporal variation is controlled 
not directly by $A$, but $N_Q$ the number of Lieb-Liniger charges forming $\Q$.  
To be sure if $A$ is large, $N_Q$ will need to be correspondingly larger in order to produce
the same minimum of temporal variation (see Eqn. \ref{strongA}).

\begin{figure}[b]
\begin{center}
\includegraphics[width=0.4\textwidth]{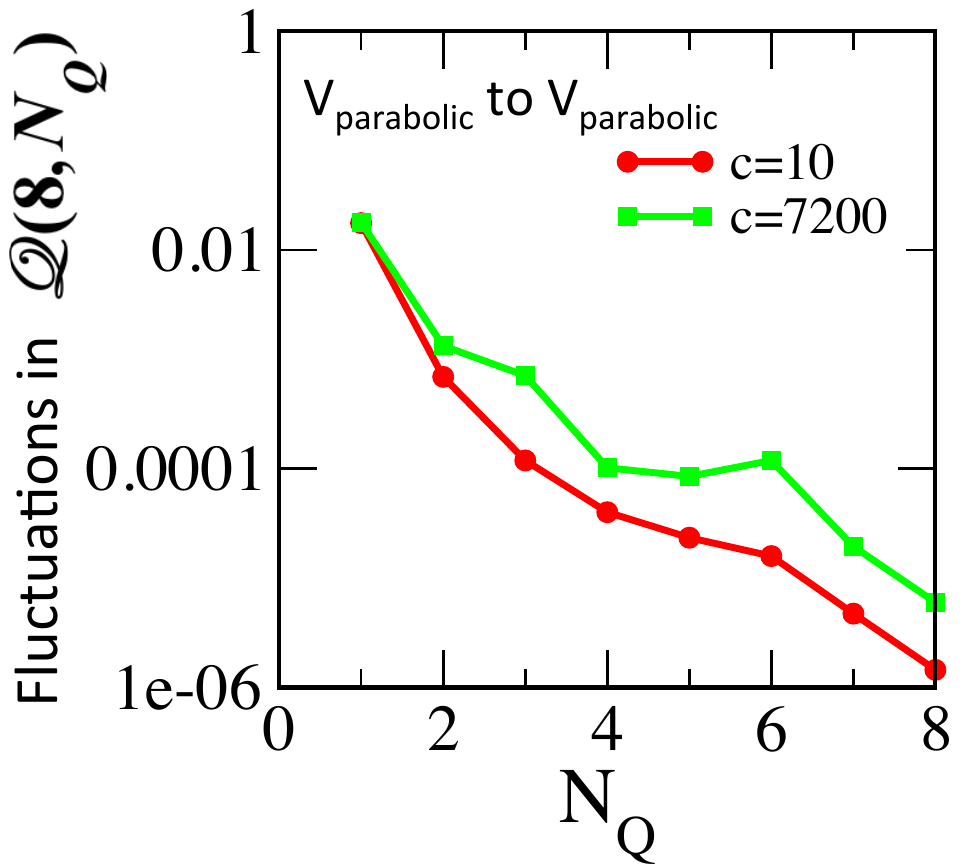}
\end{center}
\caption{\label{fluct_ptop} We show that the fluctuations in the effective charges $\Q$ constructed from a quench from
a stronger to a weaker parabolic potential, like their parabola to cosine counterparts,
die out rapidly with $N_Q$.  We consider two quenches of this type, one with
the gas at $c=7200$ and one with $c=10$.  For the $c=7200$ case, we quench from a parabolic potential 
with strength, $\omega_{0,init}$ given
by $m\omega_{0,init}^2L^2/2E_F = 6.48$ into a parabolic potential with strength $\omega_{0,fin}$ 
given by $m\omega_{0,fin}^2L^2/2E_F = 2.11$.
And for the $c=10$ case, we quench from a parabola described by $m\omega_{0,init}^2L^2/2E_F = 3.24$ into 
one given by $m\omega_{0,fin}^2L^2/2E_F = 1.06$.}
\end{figure}

In constructing these charges the nature of the potential here is important.  Our potential mixes the momenta of 
different (unperturbed) eigenstates solely through the wavevector of the cosine potential.  This is then considerably 
different than the integrability breaking considered in Refs. \cite{yur_ols,olshanii} where they considered integrability that 
respected no selection rules and correspondingly saw an extremely rapid crossover from quantum integrable to quantum chaoticity.  
However this does not mean our construction of $\Q$ does not work if the potential
induces non-trivial mixing between wavevectors.  To this end we considered preparing the system as normal in the ground
state of a parabolic potential but then instead of releasing the gas into a cosine potential, we released it into a weaker parabola.
In Fig. \ref{fluct_ptop} we show the fluctuations in $\Q (N_Q)$ as a function of the number, $N_Q$, of Lieb-Liniger charges
used to construct $\Q$.  As with the release into the cosine
potential, we are able to construct a sequence of $\Q (N_Q)$ whose temporal fluctuations
die off rapidly with increasing $N_Q$.  And although we do not show it, the off-diagonal matrix elements of these
charges fall off as rapidly as their cosine counterparts in Fig. \ref{off-diag-mes}.

\section{Experimental consequences}

\begin{figure}[t]
\begin{center}
\includegraphics[width=0.4\textwidth]{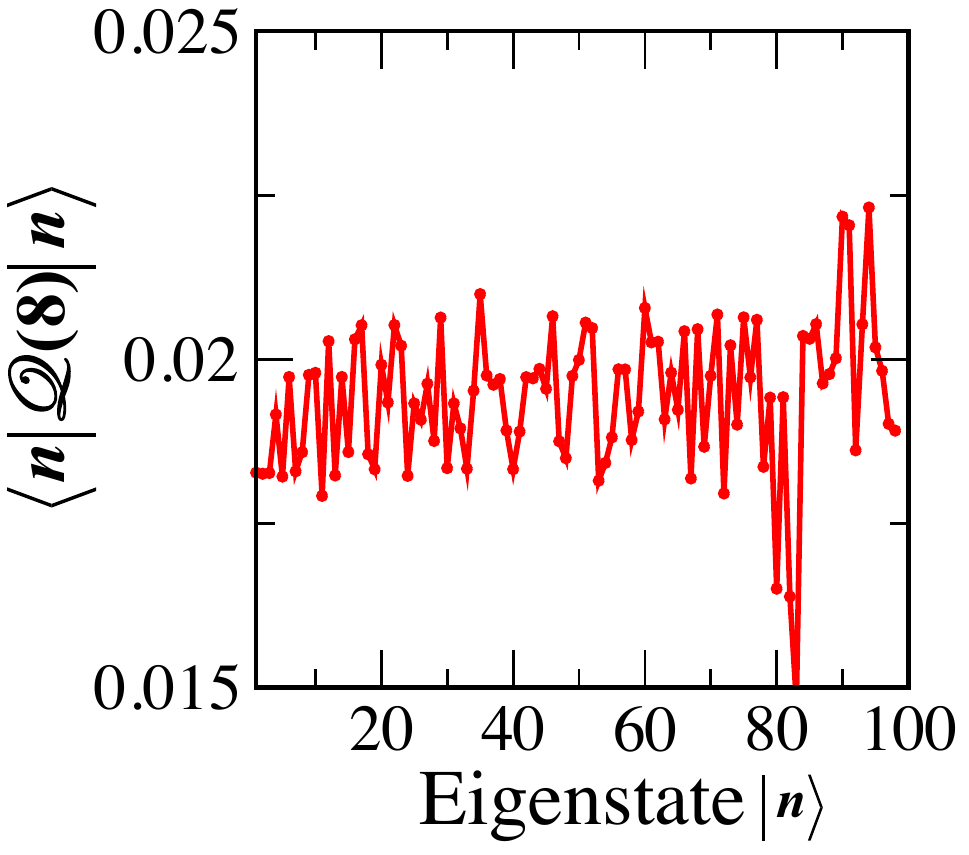}
\end{center}
\caption{\label{diag-mes}We plot the values of the diagonal matrix elements of 
$\Q (8)$ in the post-quench eigenbasis as derived for the $c=1$ quench discussed previously in
Figs. 3 and 4.}
\end{figure}

Having constructed these charges, we can ask what are the consequences of their existence.  That they take non-zero values
on the eigenstates means that the long time dynamics of the gas post-quench is going to be constrained.
In this light, we have one way to understand the ``quantum Newton's cradle'' experiment presented in Ref. \cite{weiss}.  As we
discussed in the introduction, it was argued there that the post-quench
dynamics of a gas were very slow to achieve equilibration and that this slowness was indicative of
the underlying integrability of the Lieb-Liniger model.  However, strictly speaking, the gas in this experiment
was not integrable.  The gas was confined in a one-body parabolic potential, a potential that breaks integrability \cite{note}.
Our construction of effective quasi-conserved charges in the presence of an integrability breaking one-body potential thus provides
a means to understand the slow thermalization of the gas post-quench in this experiment despite the presence of
integrability breaking.  More generally, our construction helps explain the finding of \cite{rigol_fs,rigol_fs1} where weak
integrability breaking does not lead to immediate thermalization in finite systems.

In constructing these operators, it should be stressed that the operators we construct are local
(in the sense that they are spatial integrals over operators that are defined at a single point
in space).
This follows as the effective charges, $\Q$, are constructed as linear combinations of the Lieb-Liniger charges,
which are all local quantities.  Thus we are not constructing, in effect, projection operators corresponding to eigenstates
of the post-quench gas.  Such projection operators are necessarily always present in a model regardless of its
integrability.  To demonstrate this we plot the diagonal matrix elements of the charges, $\Q$, which are linear
combinations of eight Lieb-Liniger charges and whose average
off-diagonal matrix elements are presented in Fig. \ref{diag-mes}.  We see that these matrix elements are all ${\cal O}(1)$.

If the nearly conserved quantities are governing the long time dynamics of 1D Bose gases as in Ref. \cite{weiss}, a second question that
must be asked is whether this influence is merely confined to a pre-thermalization plateau or whether it influences
the dynamics of the gas at all times.  There have been at least two constructions \cite{kollar,ess_rob} of quasi-conserved quantities
that are thought to govern pre-thermalization plateaus.   Our construction is fundamentally different inasmuch as
the quasi-conserved operators are such for all times.  This, in particular, implies that a modified form of Mazur's inequality \cite{mazur} holds.
This inequality relates the long time average of a correlation function $\lim_{t\rightarrow \infty}\langle {\cal O}(t){\cal O}(0)\rangle$
with the projection $\langle {\cal O}Q\rangle$ of the operators ${\cal O}$ onto conserved charges, $Q$.  This inequality continues
to hold with quasi-conserved charges $\Q$ but with the addition of an error term that is proportional to the size of $\Q$'s off-diagonal
matrix elements (which, in our construction, can be made arbitrarily small), something immediately clear from the proof of Mazur's inequality found
in Ref. \cite{suzuki}.  This implies that $\Q$ will control the long time limit of a host experimental observables in systems with
weak integrability breaking.  We consider this further in the next subsection.

\begin{figure*}[t]
\begin{center}
\includegraphics[width=1\textwidth]{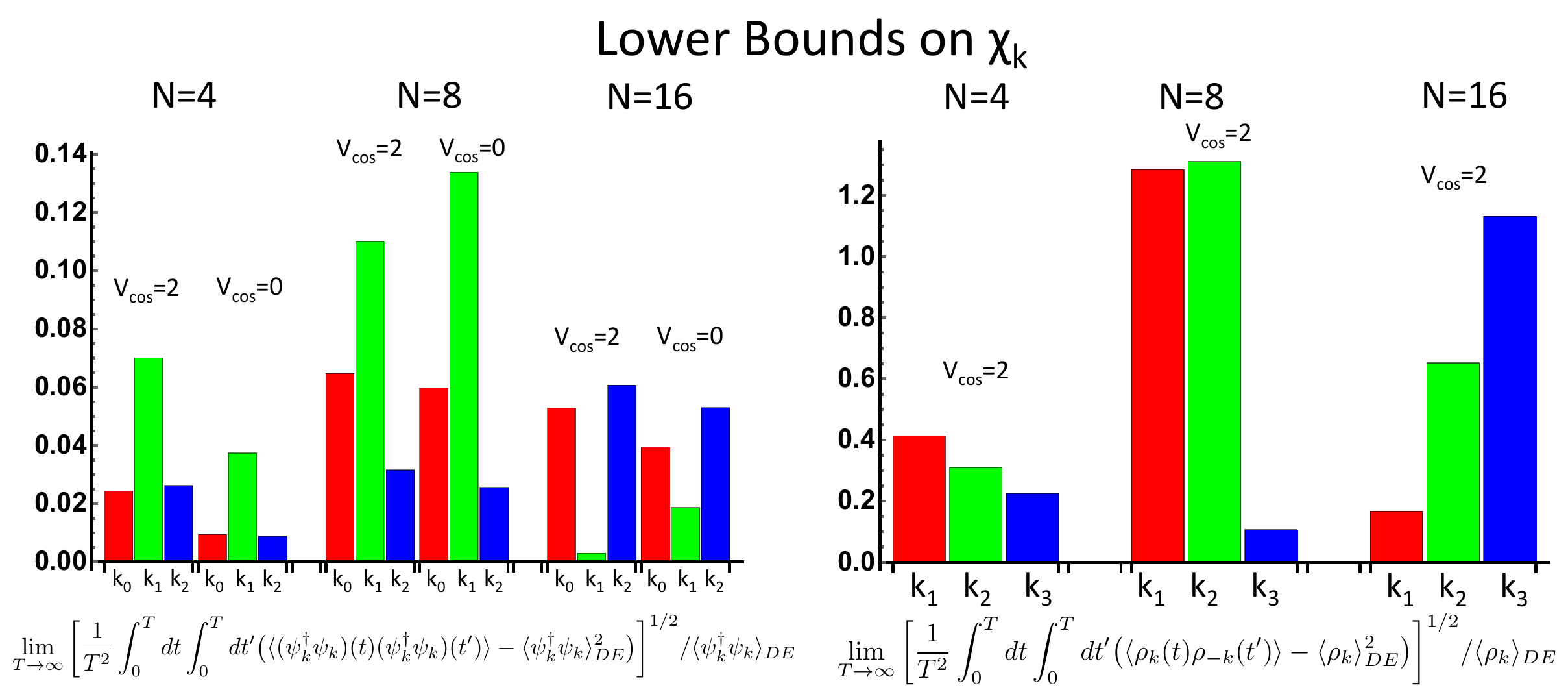}
\end{center}
\caption{\label{Mazur}
The lower bounds on $\chi_k$ due to the effective charge $\Q$ for correlators involving the MDF and the density operators.
Lefthand panel: We plot the lower bound on MDF correlations for two different quenches.  In the first (the left set of bars), we quench
into a cosine potential $V_{cos}\cos(2\pi x/L)$ of amplitude $V_{cos}=0.26E_F$.  In the second (the right set of bars) we quench into a flat potential, 
i.e. $V_{cos}=0$ with the post-quench Hamiltonian then integrable.  We present $\chi_k$ for three different system sizes $N=L=4,8,$ and $16$ 
and three different values of $k$, $k_n = 2\pi n/L$, $n=0,1,$ and $2$. The initial state of the quench is given by the 
ground state of a gas in a parabolic potential of strength $\omega = 2.4/N$.  Righthand panel: We similarly plot the lower bound on 
density correlations.  Here we only consider the case of quenching into $V_{cos}=2$ as $\chi_k$ for the density operator is identically 
zero in the absence of the breaking of translational invariance.  We again compute the lower bound at three different system sizes and 
three different wavevectors $k_1,k_2,$ and $k_3$.  In both cases we see no obvious dependence on system size.  We believe that the fluctuations seen between different system sizes results from the particular construction of $\Q$ at any given system size.  We construct
$\Q$ to minimize time fluctuations of a particular initial condition rather than construct it to maximize its overlap with a particular
observable as was done in Ref. \cite{olshaniiprl}.}
\end{figure*}

\subsection{$\Q$ and Mazur's inequality}

To understand Mazur's inequality \cite{prosen_mazur} in the context of our effective charges, we adapt the argument
presented in Ref. \cite{suzuki} establishing this inequality in the context of thermal correlation functions.  
To this end, we consider the following connected correlation function:
\begin{eqnarray}
\chi_k &=& \lim_{T\rightarrow \infty}\frac{1}{\langle M_k\rangle_{DE}}\bigg[\frac{1}{T^2} \int^T_0 dt dt_0\cr\cr
&& \hskip 0in \big(\langle M_k(t+t_0) M_k(t_0)\rangle -\langle M_k\rangle_{DE}^2\big)\bigg]^{1/2};\cr\cr
\langle M_k \rangle_{DE} &=& \lim_{T\rightarrow\infty}\frac{1}{T}\int^T_0 dt \langle M_k(t)\rangle.
\end{eqnarray}
For the case at hand, the most relevant operator, $M_k$, to consider
will be either the k-th Fourier component of the momentum distribution
function (MDF) operator, i.e., 
$$
M_k(t) = \psi^\dagger_k\psi_k ,
$$
or the density operator:
$$
M_k(t) = \frac{1}{L}\sum_q \psi^\dagger_{k+q}\psi_k ,
$$
where $\psi^\dagger_k$ is the k-th Fourier component of the Bose field.
Here we are averaging over both $t$ and $t_0$ in order to remove any dependence on the waiting time, $t_0$.  
We have defined $\chi_k$ so that correlations are measured in units of $M_k$ computed in the long time limit, i.e.
in the diagonal ensemble.
We evaluate these correlation functions $\langle \cdots \rangle$ with respect to the initial condition of
the gas in the ground state of a parabolic trap, $|i\rangle =|\psi_{GS,para}\rangle$.
$\chi_k$ is non-zero only if there are correlations present in $M_k$ that survive the $t \rightarrow \infty$ limit, i.e.
\begin{eqnarray}
\lim_{t\rightarrow \infty}\langle i|M_k(t+t_0) M_k(t_0)|i\rangle \neq &&\cr\cr 
&& \hskip -1.5in \lim_{t\rightarrow \infty}\langle i| M_k(t+t_0)|i\rangle\langle i|M_k(t_0)|i\rangle.  
\end{eqnarray}
The presence of similar long time correlations are precisely what guarantees a finite Drude weight in transport in integrable
systems \cite{zotos}. 

We demonstrate in Appendix C that a lower bound can be put on $\chi_k$ involving our effective charge $\Q$ of the form
\begin{eqnarray}
\chi_k \ge \frac{\langle i| M_k \Q_{diag}|i\rangle^2}{\langle i|\Q_{diag}^2|i\rangle} ,
\end{eqnarray}
where $\Q_{diag}$ is the diagonal part of the effective charge $\Q$.  If our initial condition state 
$|i\rangle =|\psi_{GS,para\rangle}$ is confined to the low energy Hilbert space where 
$\Q_{diag}$ and $\Q$ differ by off-diagonal matrix elements of size ${\cal O}(\delta)$, we can rewrite this inequality as 
\begin{eqnarray}
\chi_k \ge \frac{\langle i|  M_k \Q |i\rangle^2}{\langle i| \Q^2 |i\rangle} + {\cal O}(\delta),
\end{eqnarray}
as claimed at the end of the last subsection.  

We now show that this lower bound arising from $\Q$ on $\chi_k$ is in fact finite.  In Fig. \ref{Mazur} we plot this lower bound for 
both correlations involving the MDF operator and the density operator.  We study this lower bound at three different system sizes
and three different wavevectors.  We see in all cases this lower bound is appreciable.  For the MDF, the lower bound on $\chi_k$ is such
that the correlations in this quantity are at least roughly at the $10\%$ level.  To determine whether this is significant, we compute a similar
lower bound for a quench where we release the gas into a flat potential (i.e. a quench for which $\Q$ is an exact conserved quantity).  We
find values for the lower bound that are comparable to the quench into the cosine potential.  
For the density operator, the lower bound for the long time correlations
is considerably larger than that for the MDF, being bounded by values of up to ${\cal O}(1)$.  We thus see that our construction of $\Q$ acts to ensure that the system retains memory of its initial condition even at infinite time.


\section{Acknowledgements}
\begin{acknowledgements}
We would like to warmly thank both Marcos Rigol and Neil Robinson for
helpful discussions surrounding this work.
The research herein was supported by the
CMPMS Department, Brookhaven National Laboratory, in turn funded 
by the U.S. Department of Energy, Office of Basic Energy Sciences, 
under Contract No. DE-AC02-98CH10886 (RMK), by the National Science
Foundation under grant no. PHY 1208521 (RMK),
and by the Netherlands Organization for Scientific Research (NWO) and the Foundation for Fundamental Research on Matter (FOM) (JSC and GPB).
\end{acknowledgements}

\appendix

\section{Description of the 1D Bose Gas and its Post-quench Dynamics using an Numerical Renormalization Group}

\subsection{Application of the Numerical Renormalization Group }

Our approach to describing the dynamics 
associated to the quantum quench of the gas is to employ a numerical renormalization group \cite{NRG} that employs the 
eigenstates of the Lieb-Liniger model as a computational basis to determine the relatively
low lying eigenstates of the Bose gas in a one-body potential.   This numerical renormalization group is built upon both ideas taken from K. Wilson's
development of a numerical renormalization group used to study quantum impurity problems \cite{wilson} 
as well as Al. B. Zamolodchikov's numerical treatment
of perturbed conformal field theories \cite{YZ,YZ1}.  The use of the Lieb-Liniger basis as such a basis trades
on our ability to be able to efficiently compute matrix elements of relevant operators such as the density operator {\it exactly}.  While there are compact
determinental expressions for such matrix elements \cite{slavnov,slavnov1}, their evaluation is still a non-trivial numerical task and to this end we use
a set of computerized routines named ABACUS \cite{caux_abacus,caux_abacus1,cc}.
We have already demonstrated that we are able to perform the first step in our quench protocol: we have shown in Fig. 2 
that we can accurately compute the ground state of the gas in the parabolic trap.  
In this figure we plotted our numerical
determination (black) of the density profile of a gas with $N=14$ particles in a system of length $L=14$ with an interaction
parameter of $c=7200$ in a trap of strength $V_{\rm para}=\frac{1}m\omega_0^2x^2$ with $m\omega_0^2L^2/2E_F = 10.36$ against the density profile 
determined analytically (red) by mapping these (nearly) hardcore bosons onto free fermions.  The details of the analytic description
of the gas in its hardcore limit are found in Appendix A12.

In the second step of the quench protocol, we released the gas into a one-body cosine potential,
\begin{equation}
V_{\rm cosine}=\int dx A\cos(\frac{2\pi \n x}{L}).
\end{equation}
In order to compute the post-quench dynamics, 
we need to be able to describe not only the ground state in the cosine potential, but some large number of excited states.
In our quench protocol, we take as our initial $t=0$ state the ground 
state of the gas in the parabolic potential, $|\psi_{GS,{\rm para}}\rangle$.  If we can compute a wide range
of eigenstates in the cosine potential, both ground and excited states, $|\psi_{\alpha,{\rm cos}}\rangle$,
we can expand this initial state in terms of the post-quench basis:
\begin{equation}
|\psi_{GS,{\rm para}}\rangle= \sum_\alpha c_\alpha |\psi_{\alpha,{\rm cos}}\rangle.
\end{equation}
Of course for this expansion to be exact, we would need to know {\it all} of the eigenstates of the
gas in the cosine potential.  We will instead settle for a determination of the post-quench eigenbasis
that allows us to include enough states so that $\sum_\alpha
|c_\alpha|^2 > 0.99$.  We note that after we determine the initial values of the overlap coefficients, $c_\alpha$,
we proceed to normalize them so that their squares sum to 1.

In computing the spectrum of states in the cosine potential, we employ the variant of the NRG discussed
in Ref. \cite{NRG2}.  The NRG in its plain vanilla formulation \cite{NRG} 
can compute the spectrum of the low lying states of the gas in the
one-body potential \cite{NRG4}.  But to capture accurately an appreciable fraction of the spectrum, we need
to employ a sweeping routine \cite{NRG2} analogous to that used in the finite volume routine of the density
matrix renormalization group \cite{DMRG1,DMRG2}.

\begin{figure}[b]
\center{\includegraphics[width=0.475\textwidth]{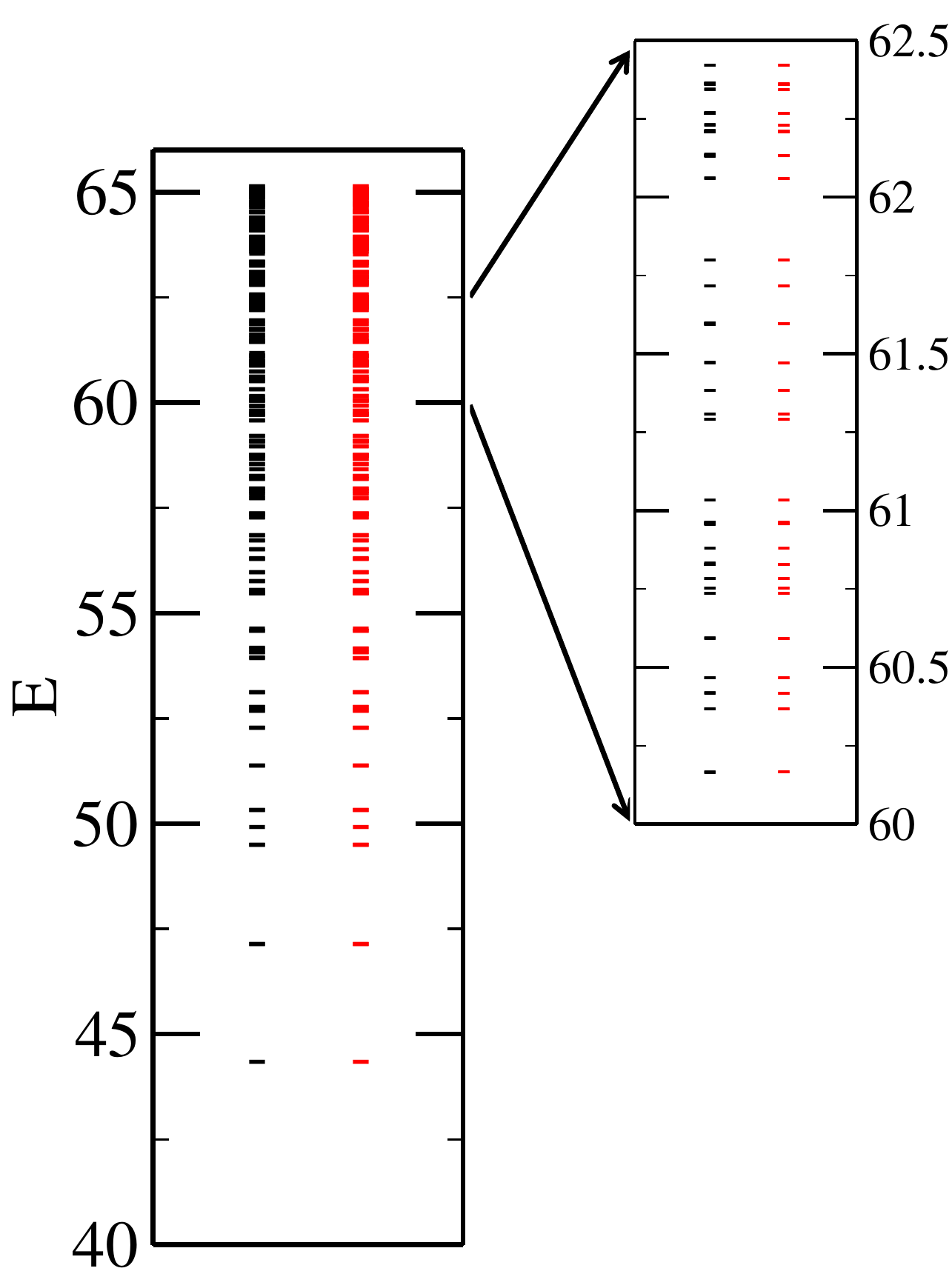}}
\caption{\label{fig:energy_spectra} A plot of the energy spectra for an N=14 gas with c=7200 in a
cosine potential of amplitude $A/E_F=0.35$ (as in Fig. 2 of the main text). 
The analytic results are given in red while in black are the corresponding numerics.
On the r.h.s. we expand a range of energy with a dense number of states so as to better exhibit agreement between
the numerics and the analytics.  We can determine the first 365 states (up to energies of $E=65$) 
with accuracy of $10^{-3}$. }
\end{figure}

In Fig. \ref{fig:energy_spectra} we present results for the spectra of an $N=L=14$ gas in the hardcore limit $c=7200$.  Here we plot in black (r.h.s.)
the numerical
determination of the first 365 energy levels of the gas in a cosine potential.  In red (l.h.s.) we plot the corresponding
analytic determination of the levels.  This analytic determination is possible by mapping the bosons to nearly free
fermions who interact with a four-body term of strength $1/c$.  Again the details of the analytics is found in Appendix A12.
The difference between the numerics and the analytics here is less than $10^{-3}$ (in absolute units).

\begin{figure}[t]
\begin{center}
\includegraphics[width=.5\textwidth]{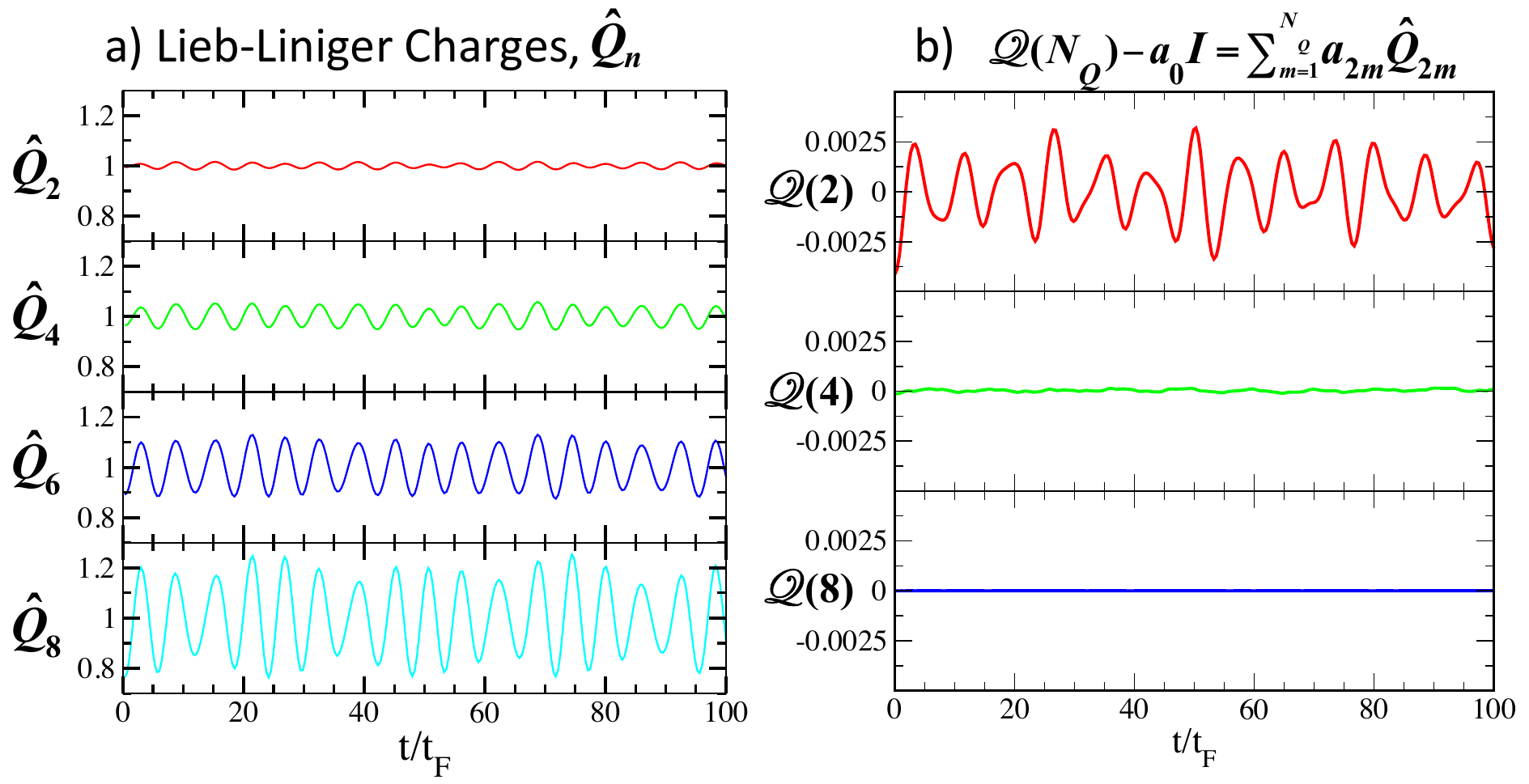} 
\end{center}
\caption{\label{fig:timeev1} a) The post-quench time evolution of the same Lieb-Liniger charges shown in Fig. 2 of the main
text but at times $t < 100 t_F$. 
b) The post-quench time evolution of the sequence of 
effective charges, $\Q(N_Q) = \sum_{m=1}^{N_Q}a_{2m}\hat Q_{2m}$, shown in Fig. 2 for $N_Q=2,4,$ and $8$ for the same range of time.} 
\end{figure}

Once we have this expansion of our initial condition $|\psi_{GS,{\rm para}}\rangle$ in terms of the eigenstates in the
cosine potential, we can readily determine the time evolution of the state post-quench:
\begin{equation}
|\psi_{GS,{\rm para}}\rangle(t) = \sum_\alpha c_\alpha e^{-iE_\alpha t}|\psi_{\alpha,{\rm cos}}\rangle.
\end{equation}
We can track time evolution of the state to a point in time determined by the accuracy by which we
can determine $E_\alpha$.  If the accuracy to which we determine $E_\alpha$ is $\delta E_\alpha$, we can
only track time evolution while $t\delta E_\alpha \ll 2\pi$ before we can no longer trust
the numerics.   Concretely, we call a state $|\psi_{\alpha,{\rm cos}}\rangle$ dephased at time $t$ if $\delta E_\alpha t > 0.01\times 2\pi$
and we conservatively will not track the time evolution beyond a point where the sum of states that are dephased have
a weight exceeding 0.01, i.e. $\sum_{\alpha \in {\rm dephased~states}} |c_\alpha|^2 > 0.01$. 
Under this criterion, we can still track the dynamics out to considerable times.
For the $N=14$ data in Fig. 2 of the main text, we can run out to times $\sim 80t_F$, while for the $N=8$
data in Figs. 3,4 and 5 of the main text, we can run considerably longer, to $t\sim 6000t_F$.  While in Fig. \ref{fig:timeev}
we present the time series for times close to this bound, we present in Fig. \ref{fig:timeev1} the time series for the same
sets of charges at shorter times, $t < 100t_F$.

With the time evolved state in hand we are able to compute the time evolution of a number of observables
and operators.
Because we use the eigenstates of the Lieb-Liniger model absent a one-body potential, $|\psi_{\alpha,LL}\rangle$,
as the computational basis of the NRG, the NRG gives any eigenstate in a one-body potential as a linear
combination of such states:
\begin{equation}\label{LLexpand}
|\psi_{{\rm one-body}}\rangle = \sum_\alpha b_\alpha |\psi_{\alpha,{\rm LL}}\rangle.
\end{equation}
Thus the dynamics of any operator whose matrix elements are known in the Lieb-Liniger basis can be determined.
As one example, we plotted in Fig. \ref{fig:timeev_rho} of the main text the time evolution post-quench 
of the density profile of the gas.

\subsubsection{Error Analysis of $\Q$ fluctuations}

One of the claims made in the text is that the effective charges $\Q$ that we construct have fluctuations
that drop exponentially with the number, $N_Q$, of Lieb-Liniger charges used in building them.  For this to be a meaningful
statement, we need to put a lower bound on the charge fluctuations arising from numerical error.

This error would arise from the dephasing errors that arise because we can only imperfectly determine the post-quench
energies.  However these errors are small.  We run out to times where only postquench
eigenstates representing $1\%$ of the weight of the initial condition have dephased 
(defined as having a phase error greater than $1\%$ of $2\pi$), i.e.
$1\%$ of the weight of the state is dephased by $1\%$.  

This might then suggest that we find a lower bound of $10^{-4}$ on the fluctuations of the effective charges.
However the off-diagonal matrix elements of the effective charges are also very small.  Thus any error due
to dephasing will be suppressed -- fluctuations in the charges are due to off-diagonal matrix elements.
So a lower bound on the error will be approximately the size of these off-diagonal matrix elements (also
on the order of $10^{-4}$) times the square root of the number of off-diagonal matrix elements
involved (square root because we assume the errors introduced by the off-diagonal matrix elements add
in the fashion of a random walk) times the error due to dephasing, so approximately $10^{-6}$ to $10^{-7}$.
This is roughly the lower bound we see on the charge fluctuations.

\subsection{Description of the gas in the cosine potential in the large $c$ limit}

In this appendix we provide a description of the hardcore limit ($c\rightarrow\infty$) of the
Lieb-Liniger model defined on a ring of length $L$ in the presence of a cosine potential:
\begin{eqnarray}
H_B &=& -\sum_{i=1}^N{1\over 2m}\frac{\partial^2}{\partial x_i^2}+c\sum_{i<j}\delta(x_i-x_j) \cr\cr
&& + A\sum_{i=1}^N\cos(\frac{2\pi \n }{L}x_i).
\end{eqnarray}
The ability to do analytics in the hardcore limit will then serve as a check on our numerical results.

For $c\gg 1$ the system can be mapped onto a system of fermions with Hamiltonian \cite{FermiBose,Tonks,Girardeau}
\begin{eqnarray}
H_F &=& -\sum_{i=1}^N{1\over 2m}\frac{\partial^2}{\partial x_i^2}-\frac{2}{m^2c}\sum_{i<j}\delta''(x_i-x_j)\cr\cr
&& + A\sum_{i=1}^N\cos(\frac{2\pi \n }{L}x_i),
\label{HF}
\end{eqnarray}
where in the dual picture we have an ultra-local interaction term of strength $1/c$.

For $c\to\infty$ the fermions are noninteracting and the physics becomes effectively one-body \cite{footnoteMathieu}.
We then must only solve the following single-body Schr\"{o}dinger equation:
\begin{equation}
-{1\over 2m}\frac{\partial^2}{\partial x^2}\psi(x)+\cos(\frac{2\pi \n }{L}x)\psi(x)=E\psi(x).
\end{equation}
This equation can be put in the standard form of the Mathieu equation,
\begin{equation}
\frac{\partial^2}{\partial z^2}\psi(z)+(a-2q \cos(2z))\psi(z)=0,
\end{equation}
if we identify
\begin{eqnarray}
z=\frac{\pi \n}{L}x; \\
q=\frac{A}{2}\left(\frac{L}{\pi \n}\right)^2 ;\\
a=E\left(\frac{L}{\pi \n}\right)^2.
\end{eqnarray}
The Mathieu equation admits Floquet-type solutions of the form
\begin{eqnarray} 
\psi_\nu^1(a,q,z)=e^{i\nu z}P(a,q,z) ;\\
\psi_\nu^2(a,q,z)=\psi_\nu^1(a,q,-z)=e^{-i\nu z}P(a,q,-z),
\end{eqnarray}
where $P(a,q,z)$ is a periodic function in $z$ of period $\pi$ 
(the same periodicity of the cosine term in the Mathieu equation).
Here $\nu=\nu(a,q)$, the Mathieu characteristic exponent function, is a function of $a$ and $q$. 
If $\nu$ is integer, the second solution is not linearly independent and a new solution must be built (see \cite{Abra}). 
In the following we will be interested only in non-integer solutions.

We are able to create linear combinations of the pairs of degenerate solutions for each triplet $\{a,q,\nu\}$.
We focus on linear combinations that are even and odd in $z$:
\begin{eqnarray}\label{Mathieu1}
\psi_{+\nu}(a,q,z)=\frac{\psi_\nu^1(a,q,z)+\psi_\nu^2(a,q,z)}{2}; \\
\psi_{-\nu}(a,q,z)=\frac{\psi_\nu^1(a,q,z)-\psi_\nu^2(a,q,z)}{2i}.
\end{eqnarray}
The final step is to construct linear combinations of these solutions that satisfy the boundary conditions. 
This step amounts to the quantization of the values of $a$, i.e. the energy, and so $\nu$.
For $N$ even, we need to impose {\it anti-periodic} boundary conditions on the single particle solutions
\begin{equation}
\psi(x+L)=-\psi(x).
\end{equation}
This will still lead the eventual N-body wavefunction to be periodic and corresponds to the use of half-integer
quantum numbers in constructing the solutions of the Bethe ansatz equations for $N$ even.
To satisfy these boundary conditions we 
choose $\nu$ to be
\begin{equation}\label{Mathieu2}
\nu=\frac{2n-1}{\n}, \phantom{oooo} n=1,2\ldots ~.
\end{equation}
It is interesting to notice that for large enough $n$, the energy $a$ coming from the two Mathieu characteristic functions corresponding 
to $\psi_{-\nu}$ and $\psi_{+\nu}$
behaves as $a \sim n^2$, as would be expected when the kinetic energy of the state greatly exceeds its potential energy.

Multi-particle states are then constructed from these single particle solutions according to 
Pauli's exclusion principle, remembering that there are two available 
states for each energy eigenvalue ($\nu$ and $-\nu$).
In comparing to the analytic solutions of the gas in the cosine potential, we perform our numerics not at $c=\infty$ but at a large finite value
of c ($c=7200$).  We thus consider perturbative corrections in $1/c$ to the hardcore limit. 
As the $1/c$ correction to the Hamiltonian,
\begin{equation}
\delta H_F = -\frac{2}{m^2 c}\int dx dx' V(x-x') \psi^\dagger(x)\psi^\dagger(x')\psi(x')\psi(x),
\end{equation}
treats two particles at a time, we can compute the correction in energy for the two-particle case and for the N-particle case simply 
add the $\binom{N}{2}$ contributions coming from all possible particles pairs.

\section{Analytic Construction of Charges in the Hardcore Limit}

\subsection{General Discussion of Analytic Construction}

We have shown that we can construct numerically quasi-conserved quantities formed as linear combinations of
Lieb-Liniger charges where the quality of the conservation is controlled by the number of charges in
the combination.  But while we have a concrete numerical construction of these new quasi-charges, we have only 
minimal analytic understanding why such charges exist.
Is this happenstance or can we provide something more solid?  The answer is that we
can, the aim of this appendix.

The basic idea behind this is to show that we can systematically construct charges of the form $\Q =\sum_i a_i \hat Q_i$ that zero out
low lying matrix elements that would otherwise lead them to have a non-trivial time dependence.  
That $\Q$ has a time-dependence at all is due to the one-body potential, $V(x)$, in the post-quench Hamiltonian:
\begin{eqnarray}
H_{\rm post-quench} &=& H_{\rm LL} + \V \cr\cr
\V &=& \int^L_0 dx V(x) \hat\rho (x) ,
\end{eqnarray}
where for us $V(x)=A\cos (2\pi \n x/L)$.  
We can rewrite this term in terms of the Fourier components of $V(x)$
and the density operator $\hat\rho(x)$:
\begin{eqnarray}
 \int^L_0 dx V(x) \hat\rho (x) &=& \sum_k V_k \hat\rho_k \cr\cr 
&=& \frac{1}{2}(\hat\rho_{k_\n}+\hat\rho_{-k_\n})
\end{eqnarray}
where $\hat\rho_{k_n} = \sum_q \psi^\dagger_{q+k_n}\psi_q$ with $k_n = 2\pi n/L$.  

The time dependence of $\Q (t)$ can be written as a power series in time $t$ 
via the Baker-Campbell-Hausdorff formula:
\begin{eqnarray}\label{BCH}
\Q (t) &=& e^{i\V t}\Q e^{-i\V t} \cr\cr
&=& \Q \!+\! it C_1 \!+\! \frac{(it)^2}{2!}C_2 + \frac{(it)^3}{3!}C_3 + \cdots\cr\cr
&&\hskip -.85in C_1 = [\sum_k V_k \hat\rho_k,\Q]; ~~~
C_{n\geq 2} = [\sum_k V_k \hat\rho_k,C_{n-1}].
\end{eqnarray}
What we now will argue is that we can systematically zero out {\it all} low energy matrix elements (below
some designated cutoff) of the first term involving the commutator of the one-body potential with $\Q (t)$.  
This results in a charge $\Q$ which has a $t^2$ (and higher) time dependence on the low energy Hilbert space.
However this higher order dependence is only nominal.  What we observe is that for a cosine one-body potential, zeroing
out first-order matrix elements also zeros out a large number of matrix elements from higher order commutators
that arise from one particle-hole processes.  To keep things tractable in this construction we will only
explicitly consider the $c=\infty$ limit where there are no more than one particle-hole processes.

To understand why higher orders remain zeroed out, we first need to describe the Hilbert space as spanned by the Lieb-Liniger eigenstates in a bit more 
detail.  An eigenstate of the Lieb-Liniger model is described by $N$-rapidities, $\lambda_i$,
\begin{eqnarray}\label{ll1}
|s\rangle &=& |\lambda_1,\cdots,\lambda_N\rangle = |I_1,\cdots,I_N\rangle,
\end{eqnarray}
which in turn are determined by $N$-integers (or half-integers) via the Bethe ansatz equations:
\begin{eqnarray}\label{ll2}
2\pi I_i  &=& L\lambda_i + \sum_{j\neq_i}\phi(\lambda_i-\lambda_j);\cr\cr
\phi(\lambda) &=& 2\tan^{-1}(\frac{\lambda}{c}).
\end{eqnarray}
We use the notion of these quantum numbers both to delineate the zeroed-out portion of the Hilbert space as well 
as to describe how it changes under higher order processes.

Let us now construct the effective charge 
$$\Q = \sum_{i=1}^{\Qm} a_i \hat Q_i ,$$ 
by defining it to have the following property: if the integers characterizing
$|s\rangle$ and $|s'\rangle$ are all such that
\begin{equation}\label{condition}
|I_i|, |I_i'| \leq N_{max},
\end{equation}
then the following matrix element vanishes:
\begin{equation}\label{C1zero}
\langle s|[\sum_k V_k \hat\rho_k,\Q]|s'\rangle =\langle s|C_1,\Q]|s'\rangle  = 0.
\end{equation}
This condition amounts to insisting that
\begin{eqnarray}\label{C1zero1}
(\Q(s')-\Q(s))\langle s|\V|s'\rangle  &=& 0,
\end{eqnarray}
where $\Q (s)$ is the action of $\Q$ on the state $|s\rangle$, i.e. $\Q |s\rangle = \Q(s)|s\rangle$.

Provided we are willing to include enough Lieb-Liniger charges in $\Q$ (i.e. choose $\Qm$ large enough) we
can always find a $\Q$ satisfying Eqn. (\ref{C1zero1}) as the collection of constraints in Eqn. (\ref{C1zero1}) form a set of homogenous
linear equations:
\begin{eqnarray}\label{conditions}
\sum_{i=1}^\Qm a_i (Q_i(s') - Q_i(s)) &=& 0, \cr\cr
&& \hskip -1.5in ~{\rm for~all} ~|s\rangle, |s'\rangle ~{\rm satisfying~ Eqn.~(\ref{condition})} .
\end{eqnarray}
The number of charges, $N_Q$ we need to include to be able to find a non-trivial solution behaves as  
$N_Q = N_{max}+2$, a number that is effectively proportional to the log of the size of Hilbert space.

\begin{figure}[t]
\center{\includegraphics[width=0.45\textwidth]{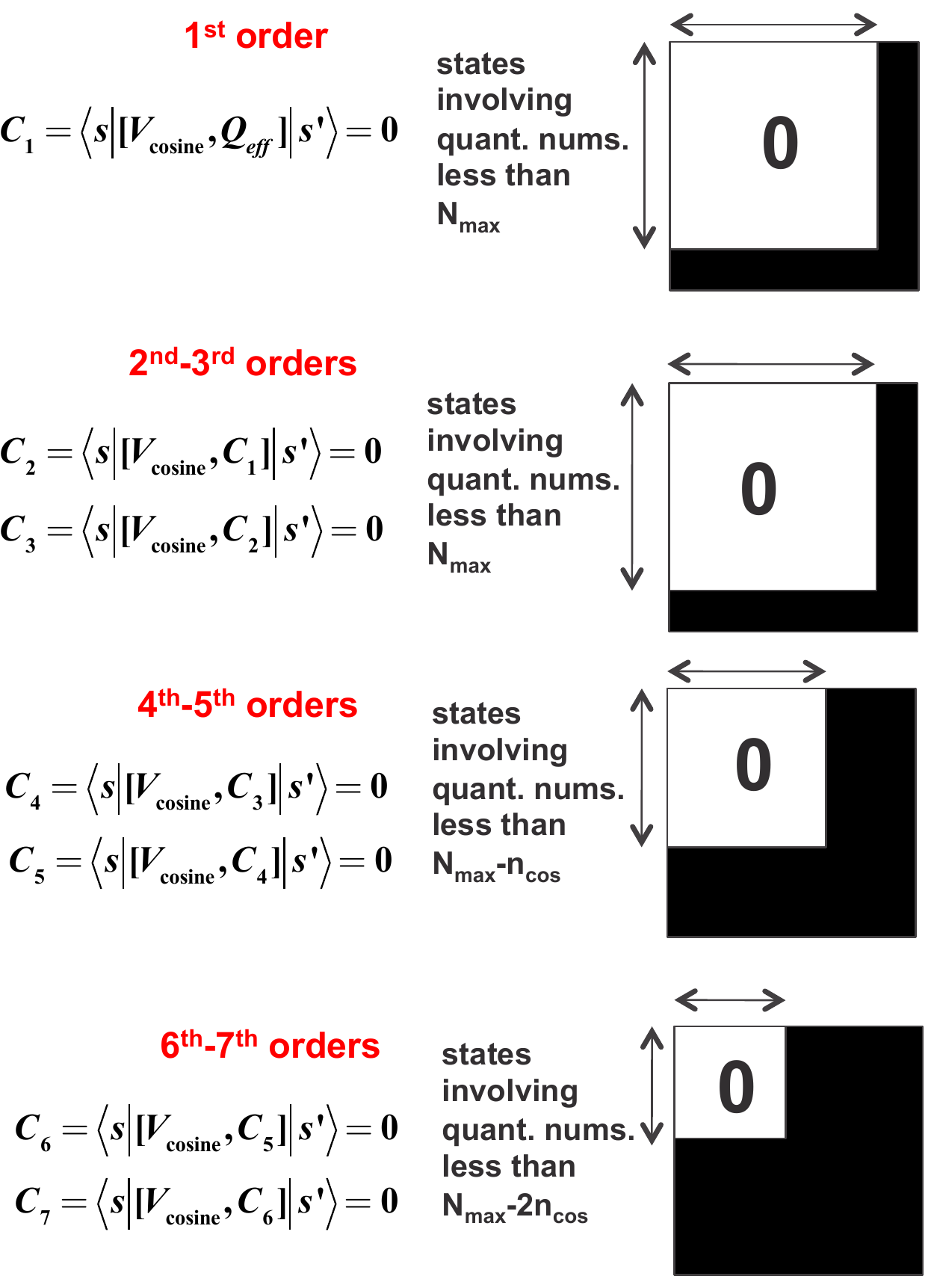}}
\caption{\label{zero_out_c_inf} Here for $c=\infty$ we illustrate how the zeroed matrix elements of $C_1$ on a low-energy block of the Hilbert space become successively
non-zero with increasing order of the higher order commutators, $C_{n>1}$.}
\end{figure}

So we now suppose we have constructed a $\Q$ where a block of states of its commutator with $\V$
have been zeroed out -- see the top square in Fig. \ref{zero_out_c_inf} for a graphical representation of this.
But now how does this zero block fare when we consider higher order commutators,
\begin{equation}
\langle s|[\V,C_l]|s'\rangle
\end{equation}
that appear in the Taylor series of $\Q (t)$.  
Roughly speaking, this block does not immediately disappear at higher order, but rather only shrinks linearly with the order of the commutator.  
The $l+1$-th order commutator, $\langle s|V^{1,p-h}_{\rm cosine},C^{1,p-h}_l]|s'\rangle$, will have non-zero matrix elements between two
states, $|s\rangle$ and $|s'\rangle$, provided that their quantum numbers satisfy,
\begin{equation}
|I_i|, |I_i'| \leq N_{max}-(l-1)n_{\rm cos}.
\end{equation}
Thus for every order in the perturbative expansion, we shrink the block of zero matrix elements
by $n_{\rm cos}$.

We can see this simply for matrix elements of the second order commutator.  Suppose then that $|s\rangle$ and $|s'\rangle$ are states
whose quantum numbers, $\{I_i\}$ and $\{I_i'\}$, satisfy
\begin{equation}\label{qnum}
|I_i|, |I_i'| \leq {N_{max}-n_{\rm cos}}.
\end{equation}
Then the matrix element $\langle s|C^{1,p-h}_2|s'\rangle$ equals
\begin{equation}
\langle s|C^{1,p-h}_2|s'\rangle = \langle s|\Voph C_1 - C_1 \Voph|s\rangle.
\end{equation}
Now the action of $\Voph$ on $|s\rangle$ is to give a state $\Voph|s\rangle = |\tilde I_1,\cdots, \tilde I_N\rangle$
whose quantum numbers must satisfy (using that the action of $\Voph$ is to change one quantum number by $\pm n_{\rm cos}$)
\begin{equation}
|I_i|, |I_i'| \leq N_{max}
\end{equation}
But by construction the matrix elements of $C_1$ between such a state and $|s\rangle$ are zero.    Hence $C_2$ has
a reduced block of zeros.   This continues on to higher order in
an inductive fashion.  The result is a shrinking block of matrix elements as pictured in Fig. \ref{zero_out_c_inf}.

\begin{figure*}[t]
\begin{center}
\includegraphics[width=\textwidth]{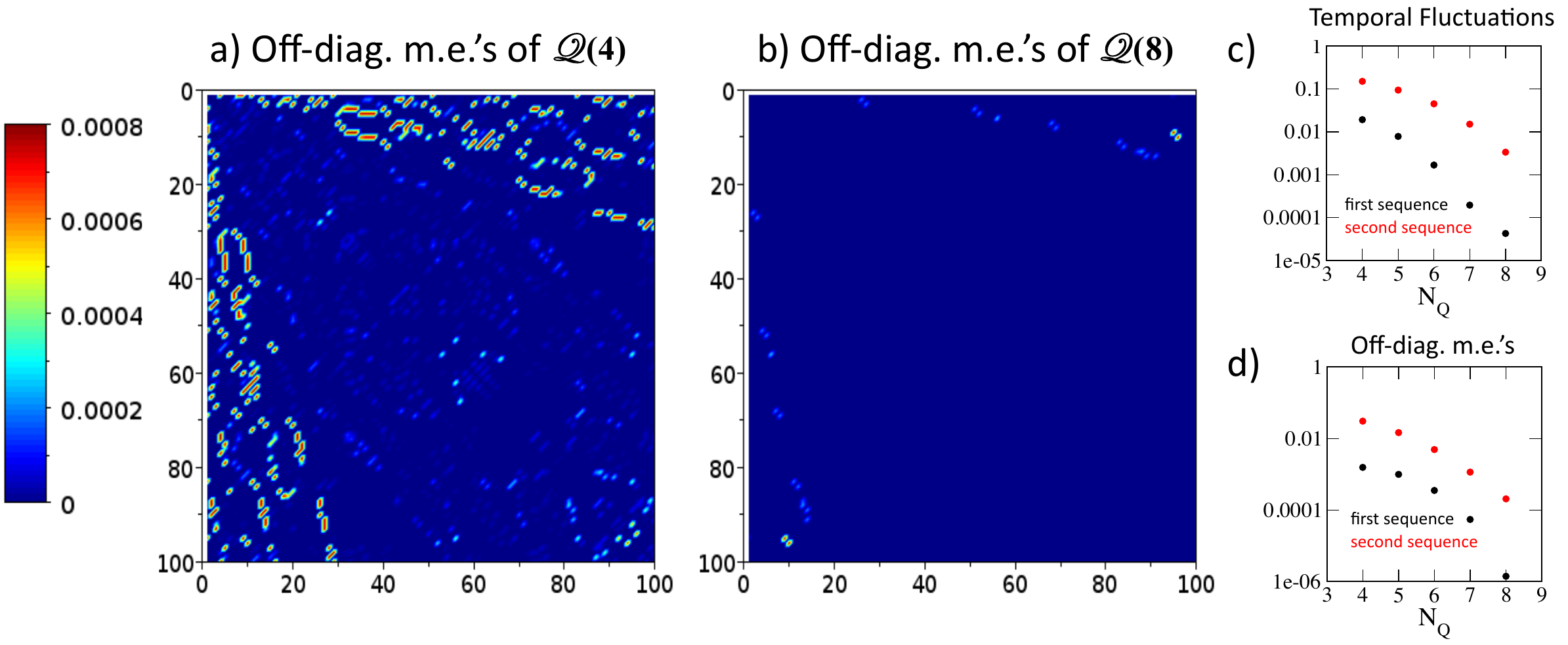}
\end{center}
\caption{\label{c_inf_anal} The magnitude of the off-diagonal matrix
  elements of an effective charge constructed from a) four Lieb-Liniger charges, $\Q (4)$, and b)
from eight Lieb-Liniger charges, $\Q (8)$.  c) The size of the post-quench temporal fluctuations of the effective 
charges $\Q (N_Q)$ as a function of the number of
charges in the linear combination.  Here the quench is performed by preparing an $N=8$, $c=7200$ gas in a parabolic 
potential of strength $m\omega_0^2L^2/2E_F=6.48$ and 
released into a cosine potential of strength $A=1$.  We show the size of the temporal fluctuations 
for two sequences of effective charges, the first (in black) constructed
from Lieb-Liniger charges, $Q_{2m}$, $m=1,\cdots ,8$ and the second (in red) constructed from $Q_{2m}$, $m=9,\cdots,16$.  
d) The size of the off-diagonal matrix
elements of these same two sequences of $\Q$ as a function of the number, $N_Q$, of Lieb-Liniger charges in the linear combination.}
\end{figure*}

Having outlined how we can construct quasi-conserved charges analytically, we now numerically test this quasi-conservation.
To perform this test, we construct a sequence of effective charges, $\{\Q_{N_{max}}\}$, defined by a sequence of maximal quantum numbers, $N_{max}$.
A charge $\Q_{N_{max}}$ is defined by its first order commutator, $C_1=[\V,\Q_{N_{max}}]$ having no non-zero matrix elements involving any two states $|s\rangle, |s'\rangle$ whose quantum numbers are less than or equal to $N_{max}$.  Such a charge will have to satisfy a number of constraints of the type
found in Eqn. (\ref{conditions}).  If there are $M$ independent constraints for a given $N_{max}$, we then form $\Q_{N_{max}}$ as a linear combination
of $N_Q(N_{max})=M+1$ Lieb-Liniger charges, specifically
\begin{eqnarray}\label{analQ}
\Q_{N_{max}} &=& a_0+ \sum^{N_Q(N_{max})}_{i=1}a_i \frac{\hat Q_{2i}}{\langle \hat Q_{2i}\rangle_{\rm av}};\cr\cr
1 &=& \sum_{i=1}^{N_Q(N_{max})}|a_i|^2;\cr\cr
\langle \hat Q_i \rangle_{\rm av} &=& \frac{1}{T} \int^T_0 dt \langle \hat Q_i (t)\rangle,
\end{eqnarray}
i.e. we include the minimal number of Lieb-Liniger charges, $\hat Q_i$, so that the null space of the set of linear equations 
in Eqn. (\ref{conditions}) has dimension 1. 
As in the main body of the text, we normalize the charges $\hat Q_i$, 
with respect to their time average following a particular quench (whose details are found in the caption to Fig. \ref{c_inf_anal}).

In Fig. \ref{c_inf_anal} we provide two tests of the quality of the conservation of the charge $\Q$ as a function of $N_{max}$.  In panel c)
of Fig. \ref{c_inf_anal} we consider the dynamics of $\Q_{N_{max}} (t)$ post-quench in our standard quantum quench protocol (preparing the gas in a parabolic
potential and releasing it into a cosine potential).  We see that the fluctuations in time of $\Q_{N_{max}} (t)$ post-quench decrease exponentially with $N_{max}$.  

However the real test of the quality of quasi-conservation of the sequence of $\{\Q_{N_{max}}\}$ is to be found in the size of their off-diagonal
matrix elements on the low energy post-quench Hilbert space.  To this end we display the size of these off-diagonal matrix elements
in panels a) and b) of Fig. \ref{c_inf_anal}.  There we show two intensity plots corresponding to $N_{max} = 4$, and $8$.  
We see that the charge $\Q$ built for $N_{max}=8$ has considerably smaller off-diagonal terms than does $\Q$ for $N_{max}=4$.
This is quantified in panel d) of Fig. \ref{c_inf_anal} where we plot the average magnitude of the off-diagonal matrix elements of $\Q_{N_{max}}$
as a function of $N_{max}$.  We see that it drops exponentially with
the number of charges.

\subsubsection{Equivalence of the Two Constructions of the Charges}

We have now demonstrated an analytic method to construct effective charges, $\Q$.  But what is the relationship between these and
those derived numerically from a particular quench protocol?  We show that in fact they do coincide.  In order to demonstrate this
we do the following.  We first fix $N_{max}$.  While we have argued that we only need $N_Q=N_{max}+2$ charges to
find a single non-trivial solution of the linear equations in Eqn. (\ref{conditions}), we consider these equations with
$N_Q=2N_{max}$ charges -- and so the linear equations will now have a null space of dimension $N_{max}-2$.  We then proceed
to find this nullspace.  Having done this, we compute numerically (as in the main text)
the effective charge built from $N_Q=2N_{max}$ Lieb-Liniger charges that arises from minimizing the post-quench temporal
fluctuations.  We then ask whether this charge (or more precisely the vector of its coefficients,$\{a_i\}^N_{i=1}$) lies in the null space
coming from building analytically the $\Q$'s.  We find that it does as $N_{max}$ grows.  This is summarized in 
Table \ref{table:overlap}.
In particular the column labelled ``Projection'' gives the projection of the normalized vector of coefficients $\{a_i\}^N_{i=1}$
into the null space (a value of $1$ indicates the numerical charge lies entirely in the null space).  We see that as $N_{max}$
increases, this projection increases quickly to its maximum possible value.  Thus we conclude the two methods are
yielding the same effective charge, $\Q$.
\begin{table}[ht] 
\caption{Degree to which numerical $\Q$ lies in null space of analytic $\Q$'s for a $c=7200$, $N=L=8$ gas:} 
\vskip 10pt
\centering 
\begin{tabular}{c c c c} 
\hline\hline 
$N_{max}$ & $N_Q$ & Dim. null space & Projection \\ [0.5ex] 
\hline
2 & 4 & 1 & 0.282 \\ 
3 & 6 & 2 & 0.718 \\ 
4 & 8 & 3 & 0.982 \\ 
\hline 
\end{tabular} 
\label{table:overlap}
\end{table}

\subsection{Estimating the Temporal Variation of $\Q (t)$}

In this section we estimate the quality of the conservation of the charges $\Q (t)$ that we have constructed in the previous section.
We do so for both weak and strong amplitudes of the post-quench cosine potential.

To determine the magnitude of the time variation in $\Q(t)$ following the quench, we 
first express the initial condition, $|\psi_{para}\rangle$, in terms of 
the post-quench eigenbasis $|\psi_{\alpha,cos}\rangle$:
$$
|\psi_{para}\rangle = \sum c_\alpha |\psi_{\alpha,cos}\rangle,
$$
and then in turn express $\Q(t)$ in terms of matrix elements of $\Q$ in this basis:
\begin{eqnarray}
\Q (t) &=& \sum_{\alpha\beta} c^*_\alpha c_\beta \langle \psi_{\alpha,{\rm cos}}|\Q (t)|\psi_{\beta,{\rm cos}}\rangle\cr\cr
&=& \sum_{\alpha\beta} c^*_\alpha c_\beta e^{-i(E_\beta-E_\alpha)t}\langle \psi_{\alpha,{\rm cos}}|\Q |\psi_{\beta,{\rm cos}}\rangle.
\end{eqnarray}
We have argued in Appendix A21 that the construction of $\Q(t)$ is such that the
time dependence (at least up to some order in time) of the low energy off-diagonal matrix elements of $\Q(t)$ are zeroed out.
This implies that some of the terms in the above expansion will be either zero (or at least small).  But which ones and what weight do they carry?
The matrix elements we have zeroed are not in the post-quench basis but in the Lieb-Liniger eigenbasis, the eigenbasis of the gas without a one-body potential.
To see the effects of this zeroing out, we expand $|\psi_{\alpha,cos}\rangle$ in terms of this basis:
$$
|\psi_{\alpha,cos}\rangle  = \sum_{I_1>\cdots > I_N} c_{\alpha,I_i}|I_1,\cdots,I_N\rangle,
$$
where the state $|I_1,\cdots ,I_N\rangle$ is constructed according to Eqns. (\ref{ll1}) and (\ref{ll2}).
We then in turn rewrite $\Q(t)$ in terms of matrix elements involving this Lieb-Liniger basis:
$$
\Q(t) = \sum_{{{\alpha,\beta}\atop{I_1>\cdots >I_N}} \atop {J_1>\cdots > J_N}} c^*_\alpha c_\beta c^*_{\alpha,I_i}c_{\beta,J_i}\langle I_i | \Q(t) | J_i\rangle.
$$
From our construction of $\Q$, we see that the matrix elements involving states $|I_1,\cdots ,I_N\rangle$ and $|J_1, \cdots , J_N\rangle$ with
$|I_i|,|J_i| \leq N_{max}$ will vanish (or at least be small).  Because all states are normalized, we know that
$$
1=\sum_{{{\alpha,\beta}\atop{I_1>\cdots >I_N}}\atop {J_1>\cdots > J_N}} |c^*_\alpha c_\beta c^*_{\alpha,I_i}c_{\beta,J_i}|^2 .
$$
To estimate how much of the time dependence of $\Q(t)$ has been eliminated, we want to compute the truncated sum,
$$
W_{elim} = \sum_{{{\alpha,\beta}\atop{I_1>\cdots >I_N}}\atop{ J_1>\cdots > J_N}} |c^*_\alpha c_\beta c^*_{\alpha,I_i}c_{\beta,J_i}|^2 \bigg|_{|I_i|,|J_i| \leq N_{max}}.
$$
The fluctuations in $\Q(t)$ will then go as $1-W_{elim}$.

In general, estimating $W_{elim}$ is difficult.  However we are able to do so in the limits of a weak and strong
cosine potential.  Because we are at large $c$, 
the pre- and post-quench wavefunctions of the N-particle gas can be described as Slater determinants of single particle states.
Pre-quench, these single particle states, $|\chi_n\rangle$, are associated with wavefunctions, $\chi_n(x)$, given in terms of Hermite polynomials:
\begin{equation}
\chi_n(x) = \frac{1}{\sqrt{2^n n!}}\bigg(\frac{m\omega_0}{\pi}\bigg)^{1/4}e^{-m\omega_0x^2/2}H_n(x\sqrt{m\omega_0}).
\end{equation}
Post-quench, the single particle states, $|\psi_{\nu}\rangle$ have wavefunctions given by Mathieu functions
(with $\nu=\pm (2n+1)/\n$) as discussed in Appendix A12 (see Eqns. \ref{Mathieu1} and \ref{Mathieu2}).  
The N-particle eigenstates can then be denoted by
\begin{equation}
|\psi_{para}\rangle = |\chi_{I_1};\ldots;\chi_{I_N}\rangle
\end{equation}
pre-quench and
\begin{equation}
|\psi_{\alpha,cos}\rangle = |\psi_{\nu_1};\ldots;\psi_{\nu_N}\rangle
\end{equation}
post-quench.
The overlap between pre- and post-quench eigenstates can then be written as a sum over products
of single-particle overlaps
\begin{equation}\label{coeff}
c_{\alpha} = \langle \psi_{\alpha,cos}|\psi_{para}\rangle = \sum_{P\in S_N}{\rm sign}(P)\prod^N_{j=1}\langle \chi_j|\psi_{\nu_{P_j}}\rangle.
\end{equation}
It is now that we specialize to the weak and strong cosine potential cases.

\subsubsection{Weak cosine amplitudes}

For weak amplitudes of the cosine potential, the post-quench single particle wavefunctions are approximately plane waves:
\begin{equation}
\psi_{\nu}(x) \approx \frac{1}{\sqrt{L}}e^{i\frac{\pi\n\nu x}{L}}.
\end{equation}
and the N-particle states $|\psi_{\alpha,cos}\rangle$ are approximately Lieb-Liniger eigenstates:
\begin{equation}
|\psi_{\alpha,cos}\rangle \approx |I_1,\cdots,I_N\rangle.
\end{equation}
The sum $W_{elim}$ simplifies in this case to:
\begin{eqnarray}
W_{elim} &=& \sum_{{I_1>\cdots >I_N}\atop{ J_1>\cdots > J_N}} |c_{I_1,\cdots,I_N} c_{J_1,\cdots,J_N}|^2 \bigg|_{|I_i|,|J_i| \leq N_{max}}\cr\cr
&=& \bigg(\sum_{{N_{max}\geq I_1>\cdots >I_N\geq-N_{max}}} \hskip -.6in |c_{I_1,\cdots,I_N}|^2\bigg)^2\cr\cr
&\equiv& X_{elim}^2
\end{eqnarray}
Because the single particle overlaps describing the N-particle coefficients, $c_\alpha$, in Eqn. (\ref{coeff}) are given by
\begin{equation}
\langle \chi_n|\psi_{\nu_{j}}\rangle = i^n\sqrt{\frac{2\pi}{m\omega_0 L}} \chi_n(\frac{\pi\nu_j\n}{L\sqrt{m\omega_0}}),
\end{equation}
we can reduce the sum $X_{elim}$ to
\begin{eqnarray}
X_{elim} &=& \bigg(\frac{2\pi}{\sqrt{\pi m\omega_0}L}\bigg)^N\sum_{N_{max}\geq I_1>\cdots > I_N\geq -N_{max}}\hskip -.4in 
e^{-\sum^N_{i=1} \frac{k_i^2}{m\omega_0}} \cr\cr
&& \hskip -.7in\times \sum_{P,P'}{\rm sign}(PP')\prod^{N-1}_{n=0} H_n(\frac{k_{P_i}}{\sqrt{m\omega_0}}) H_n(\frac{k_{P'_i}}{\sqrt{m\omega_0}}).
\end{eqnarray}
where $P$ and $P'$ are permutations of the integers $(I_1,\ldots ,I_N)$.  
In the above, the off-diagonal terms of the sum, $\sum_{P,P'}$, (i.e. those terms involving different permutations, $P_i \neq P_i'$),
are at most of order $e^{-2k_{max}^2/(m\omega_0)}(k^2_{max}/(m\omega_0))^{4N-4}$, and so can be ignored in comparison to the diagonal which take the form
$$
1-{\rm const.}\times e^{-k_{max}^2/(m\omega_0)}(k_{max}/(m\omega_0))^{2N-3}.
$$
Thus the leading order correction to the diagonal terms (which is what we care about in determining how 
much weight is left over as encoded by $1-W_{elim}$) is much larger than the off-diagonal terms which we henceforth ignore.
We can then rewrite $X_{elim}$ by converting the sums to integrals:
\begin{equation}
X_{elim}\!=\!\!\! \prod^{N-1}_{n=1}\frac{1}{\sqrt{m\omega_0\pi}2^nn!}\int^{k_{max}}_{-k_{max}} \!\!\!\!\!\! dk_i H^2_n(\frac{k_i}{\sqrt{m\omega_0}})e^{-\frac{k_i^2}{m\omega_0}},
\end{equation}
where $k_{max} = k_{max}(N_Q) = 2\pi N_{max}(N_Q)/L=2\pi(N_Q-2)/L$.  
This can then readily be computed to be
\begin{equation}
X_{elim} = 1 - \frac{e^{-\Lambda(N_Q)^2}}{\sqrt{\pi}}\sum^{N-1}_{n=0} \frac{2^n\Lambda(N_Q)^{2n-1}}{n!},
\end{equation}
where $\Lambda(N_Q) = k_{max}(N_Q)/\sqrt{m\omega_0}$.

We then see that $1-X^2_{elim}$ goes as an exponential in $N^2_{max}$ (and so $N_Q$), 
thus implying the fluctuations in $\Q(t)$ are suppressed exponentially in $N_Q^2$.

\subsubsection{Strong cosine amplitudes}

We now turn to the case of strong cosine amplitudes.  We will see the fluctuations are expected to die much more slowly with $N_Q$ than in the 
weak case.

In this limit we necessarily treat the N-particle post-quench wavefunctions as anti-symmetrized products 
of Mathieu functions labeled by $\{\nu_i\}$, i.e. $|\psi_{\alpha,cos}\rangle = |\nu_1,\cdots,\nu_N\rangle$ .
The overlap $c_{\alpha,I_i}$ is then given by
\begin{eqnarray}
c_{\alpha,I_i} = \sum_P {\rm sign}(P)\prod^N_{i=1}\langle \nu_i|n_{P_i}\rangle
\end{eqnarray}
where $\langle \nu_i|n_{P_i}\rangle$ is the overlap between a single particle Mathieu function associated with $\nu_i$ and the plane
wave $n_{P_i}$.  There is no closed form expression for this overlap (as far as we know).  However for the purposes
of this section we use the following approximate:
\begin{equation}
\langle \nu|n\rangle \approx \Theta (N_{\nu}-|n|) c_{\nu n}, ~~~{\rm if}~\frac{\nu \n}{2} \leq N_A,
\end{equation}
where $N_A = \frac{L \sqrt{2mA}}{2\pi}$ and 
where the coefficients, $c_{\nu n}$, satisfy $\sum_{n=-N_{\nu}}^{n=N_{\nu}} |c_{\nu n}|^2 =1$.  This estimate says that the expansion of a 
Mathieu function in terms of plane waves has only a finite number of terms, $2N_{\nu}$, provided $\nu$ is below a bound set by $N_A$.  Beyond this
bound, Mathieu functions becomes plane wave like (their kinetic energy is much greater than their potential energy) and their Fourier
expansion changes to consisting of a single plane wave. 
The coefficients $c_{\nu n}$ in this expansion oscillate between positive and negative amplitudes with (roughly) uniform amplitude.  While there
are Fourier coefficients of the Mathieu functions with modes beyond $N_{\nu}$, these coefficients are exponentially small in comparison
to those for $|n| \leq N_{\nu}$.

We now evaluate $\sum_{I_i} |c_{\alpha,I_i}|^2$:
\begin{eqnarray}
 \sum_{I_i} |c_{\alpha,I_i}|^2 &=& \sum_{I_i,P,P'} {\rm sign}(P){\rm sign}(P')\prod^N_{i=1}\langle \nu_i|n_{P_i}\rangle \langle n_{P'_i}|\nu_i\rangle\cr\cr
&\approx& \sum_{-N_{max}\leq I_1 < \cdots < I_N \leq N_{max}}\!\!\!\sum_{P} \prod^N_{i=1}|\langle \nu_i|n_{P_i}\rangle|^2\cr\cr
&\approx& \prod^N_{i=1}\sum_{I_i=-N_{max}}^{N_{max}}|\langle \nu_i|n_{P_i}\rangle|^2\cr\cr
&\approx& \prod^N_{i=1}\frac{{\rm min}(N_{\nu_i},N_{max})}{N_{\nu_i}}.
\end{eqnarray}
Here we make several approximations.  We take that only the diagonal terms in the sum $\sum_{P,P'}$ survive (i.e. those terms with $P=P'$).
This necessarily would happen if $N_{max} > N_{\nu_i}$ for all $\nu_i$, but because we are restricting the sum, this is merely an approximation.
It however should be a good one given that the matrix elements are bounded and oscillating in sign.  Finally we approximate the sum
$\sum_{|n|\leq N_{max}} |c_{\nu, n}|^2 = \frac{{\rm min}(N_{\nu_i},N_{max})}{N_{\nu_i}}$.  This is reasonable given the coefficients $c_{\nu n}$
are oscillating with roughly uniform amplitude in the range $n\in (-N_{\nu},N_{\nu})$.

We now need to consider the overlaps of $|\psi_{\alpha, cos}\rangle$ with the pre-quench groundstate, i.e. 
$c_{\alpha}=\langle \nu_1,\cdots,\nu_N|\chi_1,\cdots,\chi_N\rangle$.  As before, the square of this overlap can be written as
\begin{eqnarray}
|\langle \nu_1,\cdots,\nu_N|\chi_1,\cdots,\chi_N\rangle|^2 &=& \sum_{P,P'}{\rm sign}(P){\rm sign}(P')\cr\cr
&& \hskip -.5in \times\prod_{i=1}^N\langle \nu_{P_i}|\chi_i\rangle\langle \chi_i|\nu_{P'_i}\rangle\cr\cr
&& \hskip -.5in = \sum_{P}\prod_{i=1}^N|\langle \nu_{P_i}|\chi_i\rangle|^2,
\end{eqnarray}
where we suppose that this sum is again dominated by its diagonal terms.  This is justified (weakly) in that we will be performing
partial sums over the $\nu_i$'s that will (by orthogonality) provide a partial projection of the off diagonal ($P\neq P'$) terms.
We can approximate the single particle overlaps $|\langle \nu|\chi\rangle|^2$ as follows:
\begin{equation}
|\langle \nu|\chi\rangle|^2 \sim \Theta (N_A-\frac{\nu\n}{2})\frac{1}{2N_A} .
\end{equation}
Here we are using the fact that Mathieu functions with $|\nu| \leq 2N_A/\n$ (there are $2N_A$ of them in total) will
have an appreciable overlap with the Hermite function $\chi$ as such Mathieu functions have Fourier transforms that are
spread over a wide range of wavevectors with approximately equal weight.  Those Mathieu functions with
$|\nu| > 2N_A/\n$ are approximately plane waves with a large wavevector and as such with have exponentially small overlap
with the Hermite functions, $\chi$.  We thus approximate these overlaps as zero.

With this we can write down an expression for $X_{elim}$:
\begin{eqnarray}
X_{elim} &=& \sum_{{2N_A/\n\leq \nu_1< \cdots < \nu_N\leq 2N_A/\n}\atop {N_{max}\leq I_1< \cdots < I_N\leq N_{max}}}
|c_\alpha|^2|c_{\alpha,I_i}|^2 \cr\cr
&=& \prod^N_{i=1}\frac{1}{2N_A}\sum_{|\nu_i| \leq \frac{2N_A}{\n}}\frac{{\rm min}(N_{\nu_i},N_{max})}{N_{\nu_i}}.
\end{eqnarray}
Before we can evaluate this we need an expression for $N_{\nu}$.  With trial and error, we find such an expression to be
\begin{equation}
N_{\nu} = a + b\sqrt{\frac{\nu \n}{2}}N_A^\beta,
\end{equation}
with $a \approx 18$, $b\approx 1.2$, and $\beta \approx 1/2$.  This expression is approximately independent of system size $L$ and
$\n$. We can then finish the evaluation of $X_{elim}$ with the result
\begin{equation}
X_{elim} = \bigg[ \frac{N_{max}}{N_A}\frac{2}{b}(1-\frac{a}{bN_A}\log (1+\frac{bN_a}{a}))\bigg]^N.
\end{equation}
We see then that unless $N_{max}$ (and so $N_Q = N_{max}+2$) is approximately equal to the number of Mathieu functions which
have appreciable spread in Fourier space, $N_A$, the fluctuations of $\Q(t)$ that are eliminated are a small fraction of the whole.

\section{Development of a Mazur-like Inequality for $\Q$}

In this section we develop a Mazur bound arising from this existence of the effective charges $\Q$'s on the correlation function, $\chi_k$, 
involving an operator $M_k$ 
defined by 
\begin{eqnarray}
\chi_k &=& \lim_{T\rightarrow \infty}\bigg[\frac{1}{T^2} \int^T_0 dt dt_0 \big(\langle i| M_k(t+t_0) M_k(t_0)|i \rangle\cr\cr
&& \hskip .3in -\langle M_k\rangle_{DE}^2\big)\bigg]^{1/2}/\langle M_k\rangle_{DE};\cr\cr
\langle M_k\rangle_{DE} &=& \lim_{T\rightarrow\infty}\frac{1}{T}\int^T_0 \langle i| M_k(t)|i\rangle
\end{eqnarray}
We will suppose the initial condition state $|i\rangle = |\psi_{GS,para}\rangle$ is a superposition of post-quench
eigenstates whose energies all fall below a cutoff $\Lambda$.

Now our basic goal is to show that the existence of $\Q$'s places a lower bound on $\chi_k$.
The $\Q$'s that we have constructed take the form
\begin{eqnarray}
|\langle j |\Q |j'\rangle| = 
\begin{cases}
< \delta & ~\text{for }~ E_j, E_{j'} \le \Lambda ;\\
= {\cal O}(1) & ~\text{for}~ j=j' ;\\
= {\cal O}(1) & ~\text{for}~ E_j~\text{or}~E_{j'} > \Lambda ,\\
\end{cases}
\end{eqnarray}
where $\delta$ is a dimensionless number.

To demonstrate how $\Q$ controls the time evolution of $M_k$, we expand $M_k$ as follows:
\begin{eqnarray}
M_k &=& \alpha_k \Q + \sum_l \alpha_{kl} Q_l + M_k'.
\end{eqnarray}
Here $Q_l$ are some set of operators which are completely diagonal in the post-quench eigenbasis and $M_k'$ is a completely
off-diagonal operator (in the same eigenbasis).  $\tilde Q_l$ (not to be mistaken for the Lieb-Liniger charges) 
are such that they are orthogonal both to one another as well as $\Q$:
\begin{eqnarray}
\langle \tilde Q_l \tilde Q_{l'}\rangle &=& \delta_{ll'}\langle \tilde Q_l^2\rangle;\cr\cr
\langle \tilde Q_l \Q\rangle &=& 0.
\end{eqnarray}
Because $\Q$ is only (approximately diagonal) on the low energy Hilbert space, we will divide it into two pieces:
one diagonal, one wholly non-diagonal:
$$
\Q = \Q_{diag.} + \Q_{non-diag.}
$$

With this representation in hand, we now return and consider $\chi_k$.  We begin to evaluate it by inserting a resolution
of the identity between the two fields.  We will assume the spectrum is non-degenerate:
\begin{eqnarray}
\chi_k &=& \frac{1}{T^2} \int^T_0 dtdt_0 \sum_{jj'j''} c_j^*c_{j''} \langle j| M_k(0)|j'\rangle \cr\cr
&& \hskip .1in \times\langle j'|M_k(0)|j''\rangle e^{i(t+t_0)(E_j-E_{j'})+it_0(E_{j'}-E_{j''})}\cr\cr
&=& \sum_j |c_j|^2 |\langle j| M_k(0)|j\rangle|^2.
\end{eqnarray}
In this form, we see the off-diagonal parts of $M_k$ have been projected away:
\begin{eqnarray}
\chi_k &=& \sum_j |c_j|^2 |\langle j|\alpha_k \Q + \sum_l \alpha_{kl} \tilde Q_l|j\rangle|^2\cr\cr
&=& \sum_j |c_j|^2 \bigg[ \alpha_k^2 |\langle j| \Q_{diag.}|j\rangle|^2 \cr\cr
&& \hskip .75in +\sum_l \alpha_{kl}^2|\langle j| \tilde Q_l|j\rangle|^2\bigg],
\end{eqnarray}
where in the second line we have used the orthogonality of $\Q$ and the $\tilde Q_l$'s with one another.
As each term in the above is non-negative, we have the inequality:
\begin{eqnarray}
\chi_k \ge \alpha_k^2 \langle \Q_{diag.}^2\rangle
\end{eqnarray}
However for this to be a meaningful inequality we must show $\alpha_k$ is finite.

To compute $\alpha_k$ we consider the projection of $M_k$ against $\Q_{diag.}$:
\begin{eqnarray}
\langle M_k \Q_{diag.}\rangle &=& \alpha_k \langle \Q \Q_{diag.}\rangle + \sum_l \alpha_{kl}\langle \tilde Q_l \Q_{diag.}\rangle\cr\cr
&& \hskip .8in + \langle M_k' \Q_{diag.}\rangle \cr\cr
&=& \alpha_k \langle \Q_{diag.}^2\rangle,
\end{eqnarray}
where in the last line we have used the diagonality of $\Q_{diag.}$ and its orthogonality with the other charges, $\tilde Q_l$.
Thus $\alpha_k$ equals
\begin{eqnarray}
\alpha _k = \frac{\langle M_k \Q_{diag.}\rangle}{\langle \Q_{diag.}^2\rangle}.
\end{eqnarray}
By inserting a resolution of the identity between the fields and taking the action of $\Q_{diag.}$ on the post-quench eigenbasis to be
$$
\Q_{diag.} |j\rangle = \Q_j |j \rangle ,
$$
the above simplifies to
\begin{eqnarray}
\alpha_k = \frac{\sum_j |c_j|^2 \Q_j\langle j|M_k |j\rangle }{\sum_j |c_j|^2\Q_j^2},
\end{eqnarray}
while the lower bound on $\chi_k$ becomes
\begin{eqnarray}
\chi_k \ge \frac{\big(\sum_j |c_j|^2 \Q_j\langle j|M_k |j\rangle)^2 }{\sum_j |c_j|^2\Q_j^2}.
\end{eqnarray}


\begin{thebibliography}{10}
\bibitem{note}
Even though the parabolic potential mixes states with different momenta more strongly than a cosine potential, it
does not strongly connect the low and high energy parts of the unperturbed Lieb-Liniger eigenspace.
\bibitem{coeff_fn}
We have not made a systematic study of how the coefficients, $a_i$, forming the linear combination depend on $N_Q$, the number
of charges forming the effective charge.
Ideally one would want the $a_i$ to be more heavily weighted for $i$ small so as to build a more local effective charge.  Our miminization
procedure by which the $a_i$'s are constructed does not necessarily ensure this.  However we plan in future work to look at a weighted minimization scheme for the fixing of the $a_i$'s that take into account a desire for locality.  Nonetheless we do point out that the
charges we construct here, even without controlling explicitly for locality, do control long time dynamics as evidenced in Section IVa.
\bibitem{footnoteMathieu}
Even in the Tonk-Girardeau limit, the wave function of the bosonic and 
of the fermionic system are not the same, but are connected by 
$$
\psi_B(x_1,...x_N)=|\psi_F(x_1,...x_N)|
$$
The two wave functions coincide in one of the sectors, say $x_1 < x_2 . . . < x_N$, but differ by their symmetry with 
respect to the permutation of particles’ 
coordinates. While some observables are sensitive to this difference (for example the momentum distribution),
the density operator (which we will be interested in later on for computing the profile of the gas in the potential) 
does not permute particles, its matrix elements between any two many-body eigenstates of $H_B$ are identical to those
evaluated with the corresponding dual eigenstates of $H_F$.
\end{thebibliography}
\end{document}